\documentclass[runningheads]{llncs}
\usepackage[T1]{fontenc}
\usepackage{graphicx}
\usepackage{tikz}
\usepackage{amsmath}
\usepackage{xspace}
\usepackage{url}

\usepackage{blindtext}

\usepackage[inline]{enumitem}

\usepackage{xcolor,colortbl}
\usepackage{pifont}
\usepackage{diagbox}
\usepackage{adjustbox}
\usepackage{pgfplots}
\pgfplotsset{width=8.5cm,compat=1.9}
\usetikzlibrary{patterns}
\usetikzlibrary{positioning}
\usepackage{tabularx}
\usepackage{hhline,tabu}
\usepackage{multirow}
\usepackage{listings}
\usepackage{booktabs}
\usepackage{makecell}
\usepackage[labelfont=bf]{caption}
\usepackage{subfig}
\usepackage[binary-units]{siunitx}
\usepackage{dcolumn}
\usepackage{rotating}

\usepackage{lineno}
\usepackage{cleveref}
\usepackage{pgfplotstable}
\usepackage{paralist}
\usepackage[title]{appendix}
\newcommand{\ourname}{\textit{TALUS}\xspace}
\newcommand{\ournameplain}{TALUS\xspace}

\newcommand{\sgx}{SGX\xspace}
\newcommand{\txt}{Intel~TXT\xspace}
\newcommand{\tpm}{TPM\xspace}
\newcommand{\tpms}{TPMs\xspace}

\newcommand{\tpmlong}{Trusted Platform Module\xspace}

\usepackage[colorinlistoftodos,prependcaption,textsize=tiny]{todonotes}

\definecolor{bblue}{HTML}{4F81BD}
\definecolor{rred}{HTML}{C0504D}
\definecolor{ggreen}{HTML}{9BBB59}
\definecolor{ppurple}{HTML}{9F4C7C}

\definecolor{barcolor1}{HTML}{A6CEE3}
\definecolor{barcolor2}{HTML}{1F78B4}
\definecolor{barcolor3}{HTML}{9BBB59}

\newcommand{\etal}{et~al.~} 
\newcommand{\ie}{\textit{i.e.},\ } 
\newcommand{\eg}{e.g.,\ } 
\newcommand{\cf}{cf.\ } 

\usepackage{letltxmacro}
\usepackage{todonotes}
\usepackage{marginnote} 

\LetLtxMacro{\oldtodo}{\todo}
\renewcommand{\todo}[2][]{\oldtodo[fancyline,size=\footnotesize,#1]{#2}}
\renewcommand{\todo}[1]{\oldtodo[fancyline,size=\footnotesize]{#1}}

\newcommand{\scros}{\textbf{\textcolor{red}{\ding{56}}}\xspace}

\newcommand{\kinda}{\textbf{\textcolor{black}{\ding{74}}}\xspace}

\newcommand{\yes}{\textbf{\textcolor{blue}{\ding{52}\xspace}}\xspace}
\newcommand{\unknown}{\textbf{\textcolor{black}{\ding{89}\xspace}}\xspace}

\definecolor{Gray}{gray}{0.90}
\newcolumntype{a}{>{\columncolor{Gray}}c}

\newcolumntype{b}{>{\columncolor{Gray}}c}

\newcommand{\circledb}[1]{
	\tikz[baseline=(myanchor.base)]
	\node[circle,fill=black,inner sep=1pt] (myanchor) {
		\color{-.}
		\bfseries
		\sffamily
		\footnotesize #1
	};
      }

      \newcommand{\circledbs}[1]{
	\tikz[baseline=(myanchor.base)]
	\node[circle,fill=black,inner sep=1pt] (myanchor) {
          \color{-.}
          \bfseries
          \sffamily
          \tiny #1
	};
      }

\newcommand{\cheading}[1]{\textbf{\texttt{#1}}}

\definecolor{lightgray}{gray}{0.93}

\DeclareTextFontCommand{\red}{\color{red}}
\usetikzlibrary{trees,shapes.geometric,shapes.symbols,chains,calc,positioning,arrows,decorations.pathmorphing,decorations.pathreplacing}

\newlength\bubblesize
\definecolor{KWColor}{rgb}{0.37,0.08,0.25}
\definecolor{CommentColor}{rgb}{0.133,0.545,0.133}
\definecolor{StringColor}{rgb}{0,0.126,0.941}
\definecolor{Gray}{gray}{0.9}
\definecolor{lightgray}{gray}{0.93}
\begin{document}
\title{TALUS: Reinforcing TEE Confidentiality with Cryptographic Coprocessors (Technical Report)}
%
%
\author{Dhiman Chakraborty \and
Michael Schwarz \and
Sven Bugiel}
\authorrunning{Chakraborty et al.}
%
\institute{CISPA Helmholtz Center for Information Security, Saarbruecken, Germany \\
\email{\{dhiman.chakraborty,michael.schwarz,bugiel\}@cispa.de}}
\maketitle              
\begin{abstract}
Platforms are nowadays typically equipped with trusted execution environments (TEEs), such as Intel SGX or ARM TrustZone.
However, recent microarchitectural attacks on TEEs repeatedly broke their confidentiality guarantees, including the leakage of long-term cryptographic secrets. 
These systems are typically also equipped with a cryptographic coprocessor, such as a TPM or Google Titan.
These coprocessors offer a unique set of security features focused on safeguarding cryptographic secrets. 
Still, despite their simultaneous availability, the integration between these technologies is practically nonexistent, which prevents them from benefitting from each other’s strengths.

In this paper, we propose \ourname, a general design and a set of three main requirements for a secure symbiosis between TEEs and cryptographic coprocessors.
We implement a proof-of-concept of \ourname based on Intel SGX and a hardware TPM.
We show that with \ourname, the long-term secrets used in the SGX life cycle can be moved to the TPM.
We demonstrate that our design is robust even in the presence of transient execution attacks, preventing an entire class of attacks due to the reduced attack surface on the shared hardware. 

\end{abstract}
%
%
%
\section{Introduction}
\label{introduction}
%



The need for stronger protection of data and computations has led to the advent of \emph{secure enclaves}, CPU-provided isolated \emph{Trusted Execution Environments}~(TEE) that secure general-purpose computations. 
Prevalent technologies are Intel SGX~\cite{intelSGX,intelSGX2016,CostanD2016}, ARM TrustZone~\cite{armTZ}, or Keystone~\cite{keystone2020} and MI6~\cite{Bourgeat2019} for RISC-V.

The security offered by these \emph{secure enclaves} for code and data isolation depends on several high value cryptographic credentials (\eg Launch and Provisioning Key for Intel SGX, AMD PSP infrastructure key for AMD SEV, manufacturer root keys for ARM TrustZone).
Enclave programs, in turn, depend on credentials derived from those long-term secrets, \eg for secure storage of enclave data.
Unfortunately, enclave technology shares hardware, \eg CPU cores, between trusted and untrusted code, opening an attack surface. 
Especially for Intel SGX, this attack surface has been exploited in microarchitectural attacks~\cite{Schwarz2020How}, some of which leak confidential data from CPU buffers~\cite{Bulck2018,ridl2019,Schwarz2019,VanBulck2020LVI,Borrello2022AEPIC}. 

Our key observation is that virtually all platforms today are additionally equipped with specialized cryptographic or security-oriented coprocessors that protect cryptographic credentials, access control secure storage, or monotonically count.
For instance, Trusted Platform Modules~(TPM)~\cite{TPM2Spec} are available on effectively all desktop and server machines, and more solutions become available, such as Google's Titan, Microsoft's Cerberus, or AMD's PSP~\cite{Zhang2016}.
In contrast to general purpose application processors with security extensions for TEEs, those coprocessors have been designed for the primary goal to safeguard cryptographic credentials and secret data.
Integration between secure enclaves and cryptographic coprocessors creates a stronger security solution in which enclaves can use the complementary coprocessor features.
Concrete use-cases would benefit from this integration, \eg impeding microarchitectural attacks against enclaves based on TPM features.
Unfortunately, such an integration is currently, if it exists, very limited.
We ask the following fundamental questions:
\textit{Which security guarantees does the combination of CPU-provided TEEs with secure coprocessors provide that each of the technologies cannot provide on their own?
    What are the requirements to combine the advantages of both technologies without introducing new security problems or large performance overheads? }

To answer these questions, we introduce a hardware/software co-design, \ourname
, to combine CPU-provided TEEs with cryptographic coprocessors.
Enclave code can directly invoke the coprocessor \emph{only} via the CPU firmware and bus connections to make use of the coprocessor's facilities, such as counters or key management.
We identify three core requirements to realize our idea: a \emph{secure communication channel} between processors and coprocessors, \emph{vertical access control} to distinguish between enclave and non-enclave code, and \emph{horizontal access control} to distinguish between different enclaves.
To understand how SGX can be integrated with an on-board hardware TPM, we built a proof-of-concept integration between Intel SGX and a hardware TPM on commodity hardware.
We show that a combination of Intel SGX (emulated through KVM-SGX~\cite{kvmSGX} and QEMU-SGX~\cite{qemuSGX}) with hardware TPM is feasible with firmware changes and demonstrate through different use cases the security benefits of this symbiosis.

We show that \tpm fills a gap in the trusted-computing features of SGX that is due to a lack of replay-protected secure non-volatile memory.
Several previously published defenses for attacks against SGX provide their full strength only if such building blocks are available~\cite{Shih2017tsgx,Oleksenko2018varys,Chen2017detecting}.
Furthermore, preventing recent microarchitectural attacks against TEEs~\cite{controlledsidechannel,Bulck2018,ridl2019,Schwarz2019}, including undervolting~\cite{Pengfei2019,Murdock2019plundervolt,Kenjar2020v0ltpwn} 
is only effective if an enclave can store a persistent state to limit the number of attack attempts. 
%
%
In addition to the possibility of preventing attacks against enclaves, we demonstrate that all high-value secrets used during the lifetime of enclaves can be safely stored in the TPM without ever reaching a shared hardware element.
We can actively mitigate existing attacks and harden an enclave against potential future attacks by reducing the amount of high-value secrets stored in the enclave.
Our proof-of-concept implementation shows that the expected overhead of an average 21.6\% is amortized in typical use cases, as only rarely used operations suffer from a slowdown of several milliseconds.

In summary, we make the following contributions:
\begin{compactenum}
 \item We introduce \ourname, a hardware/software co-design to combine CPU-provided TEEs with cryptographic coprocessors.
 \item   We show that \ourname provides extended features, like rollback protected \tpm NV-storage for persistent counters to limit execution control attacks against enclaves.
 \item We demonstrate that \ourname significantly reduces the attack surface for microarchitectural attacks.
 \item We analyze \ourname for real use cases, showing that its performance overhead is amortized in many use cases while providing strong security guarantees.
\end{compactenum}



%

\section{Background}
\label{background}
%


\subsection{Intel Software Guard eXtension (SGX)}
\label{background:sgx}


\sgx is an extension to the x86 instruction set that allows a user-space process to create and manage a protected isolated memory region called an \textit{enclave} within its own address space, even protected from OS and hypervisor access~\cite{mckeen2013,intelSGXDevMan2016}.
\sgx assumes that the CPU, including its microcode, is the only trusted element in the system. 
Enclave data are stored encrypted in DRAM and unencrypted in the CPU caches and registers.
An external party can verify an enclave by (remote) attestation of the enclave code and meta-data~\cite{intelSGXRemAtt2016,Brickell2012}.

Intel supplies two infrastructure enclaves, the \textit{launch enclave}~(LE) and \textit{quoting enclave}~(QE), on which \sgx is heavily dependent.
The LE is responsible for handling and launching user-space enclaves with a token called \lstinline{EINITTOKEN} that is generated using i)~the measurement of the static content of the enclave (\lstinline{MRENCLAVE}) and ii)~the enclave-author validation (\lstinline{MRSIGNER}).
The LE requires a 128-bit \emph{Launch Key} (LK) to derive the \lstinline{EINITTOKEN}. 
The QE is designed to validate local attestation reports by enclaves generated with an asymmetric private key that a remote verifier can verify.
Both the LE and the QE are entrusted with long-term high-value cryptographic credentials. 

\subsection{\tpmlong (TPM)}
\label{background:tpm}

\tpm by the Trusted Computing Group is the most widely deployed trusted computing technology on commodity platforms used by, \eg Microsoft Windows management instrumentation, Intel Trusted eXecution Technology~(TXT)~\cite{intelTXT2021}, Microsoft Bitlocker~\cite{bitlockerTPM}, or Google Chrome~\cite{chromiumTPM}. 
A TPM contains a small non-volatile memory block, a set of platform configuration registers (PCR), an onboard processor to execute \tpm code in isolation from the other hardware, co-processing for standard cryptographic algorithms, a secure clock, and a random number generator.
TPMs can reliably report internal data to a third-party verifier, i.e., remote attestation based on a pre-installed endorsement key.
Typically, a \tpm is available as a hardware chip soldered to the mainboard, traditionally connected via the Low Pin Count (LPC) bus or on newer platforms via the SPI bus, making it only available through memory-mapped I/O (MMIO) registers protected by the chipset.
Intel also implements a firmware TPM called Platform Trust Technology (Intel-PTT)~\cite{intelPTT2014} housed inside Intel CSME~\cite{intelCSME2020}.
\section{Requirements Analysis}
\label{ReAnalSystematization}
%


In this section, we define three fundamental requirements for a secure integration of CPU-provided Trusted Execution Environments with onboard secure coprocessors: \emph{secure communication channel}, \emph{horizontal access control}, and \emph{vertical access control}.
We systematically compare how \sgx and \tpm meet those requirements and how well these two technologies can be integrated, as demonstrated later by our proof-of-concept implementation.
In Appendix~\ref{app:ReAnalSystematization}, we have extended this comparison further to other secure coprocessors and TEEs.

\noindent
\linebreak
\textbf{Communication Channel (CC).}
For a secure integration between the security coprocessor and the application processor (AP), the communication channel between them must be secured from eavesdropping even in case of physical attacks, \eg bus sniffing (\textbf{CC1}), and there should not be any dependencies on buffers vulnerable to microarchitectural attacks that can leak sensitive data transferred via the channel (\textbf{CC2}).
TPM and SGX fulfill \textbf{CC1} since \tpm and Intel CPUs support end-to-end encryption of the communication between them~\cite{TPM2Spec,Futral2013,intelTXT2021}.
However, this channel does not avoid insecure buffers, and decrypted data on the CPU side might still pass through such buffers.
As demonstrated by recent attacks, none of the TEEs, including SGX, is free of insecure buffers.
Therefore, SGX inherently fulfills \textbf{not} \textbf{CC2}, and we show in Section~\ref{systemDesign} how we overcome this limitation in combination with \tpm.

\noindent
\linebreak
\textbf{Horizontal Access Control (HC).}
TEEs can host multiple tenants.
For example, SGX supports multiple (parallel) enclaves.
Horizontal access control ensures that the AP and the coprocessor can distinguish between requests from mutually untrusted tenants inside a TEE.
For instance, one enclave should not be able to access another enclave's data within the coprocessor.
A trusted entity, such as an AP, must create access or identity tokens that can identify TEE tenants.
The tokens must be securely communicated to the coprocessor.
The coprocessor must also understand those tokens to control access to managed data and secrets.
Hardware \tpm and firmware \tpm employ extended authorization policies (EAP) that can use these access tokens for access control to TPM-managed objects, like TPM-internal storage and keys.
All AP-based TEEs can fulfill this requirement because they can uniquely identify the different enclave codes they host.
They can provide this information on calls to the coprocessor.
For example, in SGX, this would be the code measurement of the enclave by the CPU.

\noindent
\linebreak
\textbf{Vertical Access Control (VC).}
AP-based TEE technologies and the coprocessor should support access control based on different security levels (\eg application, OS, or hardware) to prevent non-enclave code from accessing enclave-owned entities in the coprocessor.
The access token to distinguish between different security levels needs to be generated and handled by a secure piece of code and be securely communicated to the coprocessor.
Hardware and firmware \tpm offer \emph{Locality} to distinguish between TPM commands originating from different security levels. Still, the locality of a command must be communicated to the TPM by the CPU or firmware.
Furthermore, SGX registers when it executes in enclave mode, but this security level is only used CPU-internally and not for Locality.

\noindent
\linebreak
\textbf{\ournameplain: Integrating Intel SGX and Hardware TPM.}
The main issue of vanilla \sgx is the lack of confidentiality- and integrity-protected tamper-resistant storage.
As we are unaware of any non-volatile memory inside a CPU, we do not see how SGX can be improved by only updating the firmware and without adding new components (like a TPM) to the TCB.
Vanilla \sgx can use PTT for certain trusted computing use cases.
However, PTT is housed inside the CSME~\cite{intelCSME2020} and connected through the DMI interface without any security around the communication channel.
Moreover, although CSME employs its own OS with its own security ring, completely segregated from the platform security, the command buffer for PTT is configured by untrusted software, such as the OS, and PTT recently suffered from access control errors~\cite{csmeATTACK2018,csmeATTACK2019} that completely undermine its security and are currently unfixable in production devices.
Additionally, secrets typically flow through the memory hierarchy on the CPU where untrusted code can run in parallel, observing side effects of the secret processing, e.g., when unsealing data from disk.
Furthermore, in \sgx, support for counters depends on the Platform Service Enclave and Intel ME, which are often not available in \sgx production deployments and have already been deprecated~\cite{intelCounter}.
Moreover, these counters can be reset by reinstalling the SGX platform software~\cite{MateticAKDSGJC17}.
As \sgx stores counters inside the BIOS flash storage, they do not persist across system resets~\cite{MateticAKDSGJC17}.
The unavailability of integrity-protected, tamper-resistant storage does not allow SGX to store a secure counter, which limits the possibility of enclaves to enforce a number of enclave executions, as exploited in interrupt-based attacks~\cite{Bulck2018}.

Based on our requirements analysis, we found that the combination of SGX with hardware TPM is highly amenable for integration and allows to fill those gaps in SGX with TPM functionality.
Due to the historical relationships between Intel CPUs and TPM, they can create an encrypted channel between them.
Additionally, SGX can identify (i.e., measure) enclave code while TPM can use this identity in its access control policies.
Therefore, our proof-of-concept implementation for our \ournameplain design is based on SGX and a \textbf{hardware} TPM.
\section{High-level \ournameplain Overview}
\label{integrationDiscussion}
%




\Cref{fig:highlevelfigure} illustrates the high-level overview of \ourname.
Our systematization (\Cref{ReAnalSystematization}) underlines the intuition that the TPM, when integrated as a coprocessor with SGX, can provide desirable features to secure enclaves, such as physically isolated processing of cryptographic secrets, a secure clock, or persistent counters.
The basic idea is to retrofit SGX with a \emph{direct} communication channel to the TPM chip \emph{without} going through the host OS.
With such a communication channel, enclaves can leverage the TPM facilities as building blocks, \eg to implement secure monotonic counters (\cf \Cref{systemDesign}).
This section provides more details on the security benefits, requirements, and challenges of integrating \sgx enclaves with a TPM.

\subsection{Threat Model}
\label{threatmodel}

The threat model for \ourname is the union of the coprocessor and enclave threat model.
Only the coprocessor (including firmware) and the processor (including microcode) are trusted.
We assume that the coprocessor does not suffer from implementation~\cite{244048} or platform integration flaws~\cite{Han2018}.
Similarly, we assume that the enclaves are not malicious~\cite{Schwarz2019Practical} and are free of classical software vulnerabilities~\cite{Weichbrodt2016,Lee2017SGXROP,VanBulck2019tale,Cloosters2020teerex}.
Microarchitectural attacks~\cite{vanbulck2020phd}, such as classical side channels and transient execution attacks, are in scope.
%
%
We allow physical attacks in line with the \tpm and \sgx specifications, \eg bus tapping, bus sniffing, or similar physical layer attacks~\cite{tpmgenie,conf/uss/Kauer07,TPMHWattack}. 
We exclude physical attacks outside of a reasonable attacker model for \sgx and consumer-grade hardware, such as bus snooping on high-speed or address buses~\cite{membuster2020}, against which \sgx also fails to defend.



\subsection{Design of \ournameplain}
\label{challengesforintegration}

Integrating a coprocessor (\eg \tpm) with a secure enclave technology, such as \sgx, poses both security~(\textbf{\texttt{SC}}) and functional~(\textbf{\texttt{FC}}) challenges.
In this section, we detail the challenges and how we design \ourname to solve these challenges.

\textbf{{SC1. Secure Communication Channel.}} 
CPU and coprocessor must exchange data securely.
Ideally, the coprocessor is physically integrated with the CPU package (e.g., similar to AMD PSP), and the communication channel is physically secured against eavesdropping.
If the coprocessor is an additional hardware element, a secure connection via the usually insecure bus is required.
For \tpm and \sgx, \tpm is connected to the CPU via the unprotected LPC or SPI bus.
Thus, \ourname relies on symmetric authenticated cryptography to establish a secure channel between the coprocessor and the CPU while ensuring confidentiality and integrity despite an untrusted OS and a physical attacker.

{\textbf{{SC2.} Authorization of Commands.}} A coprocessor, such as \tpm, is often shared between various entities on the system, such as firmware, OS, and user-space applications.
Further, the enclave technology might support multiple mutually-untrusted tenants.
Thus, the coprocessor has to manage the credentials for different enclaves (differentiated using, \eg \lstinline{MRSIGNER}, \lstinline{MRENCLAVE}, \lstinline{PRODID} and \lstinline{SVN}).
Moreover, the coprocessor is also used by non-enclave code, e.g., the OS, firmware, or user-space application.
Consequently, it is crucial to have authorization of coprocessor commands to control access to coprocessor entities (like keys or NVM) to ensure that every enclave and non-enclave code only ever has access to its own coprocessor entities.
\ourname with \sgx and \tpm ensures authorization using locality and EAP.
Authorization to \tpm entities between different actors in the system, \eg OS, third-party software, or hardware, is based on the \tpm locality.
Different enclaves running on the same system authorize via their identities through \tpm EAP (\cf \Cref{fig:highlevelfigure}).

\textbf{{SC3.} Avoiding Shared Hardware.}
It is often necessary to securely (\cheading{SC1}) send secret data, \eg session keys, to the CPU while reducing the amount of shared hardware involved in the communication.
Recent transient-execution attacks showed that a software-only attacker can read stale entries in various internal CPU buffers~\cite{Bulck2018,Schwarz2019,ridl2019,fallout2019,ragab_crosstalk_2021}.
Thus, \ourname provides strict isolation of coprocessor-released data, ensuring that data does not pass (in plaintext) through shared hardware elements with (known) vulnerabilities.
\ourname implements the entire communication using only CPU registers as storage.

\medskip
Besides those security challenges, we identify the following functional challenges (\cheading{FC}) that influence \ourname.

\textbf{{FC1.} Functionality Mapping.}
Enclave functionalities require a corresponding faithful command mapping offered by the coprocessor, \eg to generate and use keys with the same authorization policies.
%
The coprocessor driver logic for these commands can be implemented in CPU microcode~\cite{intelXuCode} without requiring hardware changes.
The microcode changes have to support only minimal amounts of ephemeral storage for policy sessions and session handles, both of which can be stored in the insecure BIOS flash.

\textbf{{FC2.} Attestation.}
Enclaves depend on attestation to convince (remote) parties that they are communicating with the intended enclave.
If the coprocessor supports attestation and management of attestation secrets, the attestation can be outsourced to the coprocessor.
Thus, attestation secrets are never stored in shared hardware.
A \tpm supports remote attestation of \tpm internal data.
However, this poses the challenge of faithfully integrating the \tpm attestation protocols with \sgx.
\ourname achieves this by extending \tpm PCR21 with a measurement of an SGX secret (\eg measurement of the QE).
PCR21 is protected using EAP to ensure that only the microcode can access it, and the PCR21 measurement is attested through \tpm-based attestation to a remote verifier. 

\textbf{{FC3.} Asynchronous Execution.}
When outsourcing cryptographic commands to a potentially much slower coprocessor, we face the problem that the coprocessor execution is asynchronous to the enclave execution. 
For example, the enclave might be interrupted before the coprocessor finishes executing an issued command by the enclave. 
Thus, \ourname ensures proper scheduling between enclave execution and coprocessor execution to handle asynchronous execution by storing secrets in the special-purpose registers and encrypting them during interrupts, preventing the register content from leaking through unprotected buffers. 
Interrupts already require a significant amount of microcode execution in the CPU, \eg SGX stores registers in the SSA and resets the register values to non-secret values. 
Hence, adding encryption is feasible in microcode.


%

\section{\ournameplain Implementation}
\label{systemDesign}
%

This section briefly introduces the implementation details of a proof-of-concept of \ourname based on \sgx and a hardware \tpm. An in-depth discussion is available in Appendix~\ref{app:systemDesign}.
We show the functionality and all the security guarantees using the Intel SGX emulator~\cite{intelVirtSGX} and a hardware \tpm, allowing us to implement the entire design of \ourname.
For the performance evaluation, we instead use a hardware SGX enclave in combination with the same hardware \tpm, with the limitation that the communication channel is not protected against a malicious OS.
All evaluations are performed on an Intel i7-7820X running Ubuntu 16.04.04 with kernel 5.0.0.
As the TPM, we use an Infineon SLB 9670 that supports TPM 2.0 (HTPM). The size of the enclave used for performance evaluation is \SI{52}{\kilo\byte}.

%

\subsection{Connecting SGX and TPM}
\label{harpocrates}


\textbf{Channel between \sgx and \tpm (\cheading{SC1}).} \label{channelBetweenSGXandTPM}
Typically, the OS provides the \tpm as an MMIO device to the system and user-space software.
However, \ourname \textit{cannot} rely on the untrusted OS for communication. 
For our proof-of-concept implementation, we rely on the end-to-end encrypted programmed I/O channel between the CPU and the hardware \tpm.
To prevent untrusted system software from interfering with the channel, we distinguish between MMIO and DMA requests. 
The channel is controlled by \txt using an access control mechanism called \textit{Locality} offered by the \tpm through \textit{\tpm Locality Address Mapping}~\cite{intelTXT2021}.
\tpm localities indicate the source of the command within the platform.
Locality 0 is full public access, locality 1 is the OS, and higher localities (up to locality 4) correspond to the highest privilege levels, \ie hardware and microcode, including \sgx.
In \ourname, localities ensure the vertical access control to the \tpm (\eg software, OS), while command authorization (\cf \Cref{sec:integration}) ensures the horizontal access control (\ie different enclaves).

%
The channel directly stores data in the CPU registers.
Cole and Prakash~\cite{Cole2020simplex} showed that, in addition to general-purpose registers, sensitive data can also be stored in the Intel MPX \texttt{bnd} registers.
As Linux or GCC no longer supports Intel MPX~\cite{Phoronix2020MPX}, these registers can be used by an enclave without conflicting with any other existing software.

\textbf{Interrupt Handling (\cheading{FC3}).}
On an interrupt, \sgx performs an \textit{Asynchronous Enclave Exit} (\lstinline{AEX}) to save the enclave execution state in the \textit{State Save Area} (SSA) before invoking the OS exception handling.
Although architecturally secure, RIDL~\cite{ridl2019}, ZombieLoad~\cite{Schwarz2019}, and \AE PIC~\cite{Borrello2022AEPIC} showed that storing registers in the SSA leaves copies of the values in internal CPU buffers from where they can be leaked.
Forcing \sgx to dump registers to the SSA is always possible, as an attacker can inject interrupts at any time during enclave execution~\cite{sgxstep2017}.

\ourname does not allow the registers (\lstinline{BND0}-\lstinline{BND3}) holding potentially secret data to be saved directly to the SSA.
In our proof-of-concept implementation, we encrypt the registers on \lstinline{EEXIT}, \lstinline{EREMOVE}, or \lstinline{AEX} before storing them.
We use AES in counter mode, with the SGX sealing key as the encryption key and the number of asynchronous exits as the counter.
Using the number of asynchronous exits as a counter has the advantage that an attacker has only one shot at leaking the (encrypted) secret, and the attacker cannot even detect if the secret has changed between two interrupts~\cite{Li2021cipherleaks}.

As computations with secrets often require multiple general-purpose or SIMD registers~\cite{Muller2011tresor,Garmany2013prime}, it is also beneficial to prevent other registers from spilling secrets into the SSA.
Similarly to protecting enclaves from traditional side-channel attacks, we see that responsibility with the enclave developer. 
Without \ourname, a developer cannot write code so that secrets are not leaked through transient-execution attacks.
If TSX is available, it is possible to protect intermediate results from spilling into the architectural domain by relying on a compiler extension~\cite{Gruss2017Cloak}.
However, since TSX is deprecated, transient execution can be used as a (less efficient) alternative, as shown in recent work~\cite{Wampler2019exspectre,Schwarz2019,Zhang2020exploring}.

\subsection{Porting SGX Functionality to TPM}
\label{sec:integration}

In this section, we demonstrate that \sgx functionality can be mapped to the \tpm using command authorization.


\begin{figure}[t]%
	\centering
	\subfloat[\tpm communication from user space enclaves for \sgx operations.]{{\includegraphics[width=5cm]{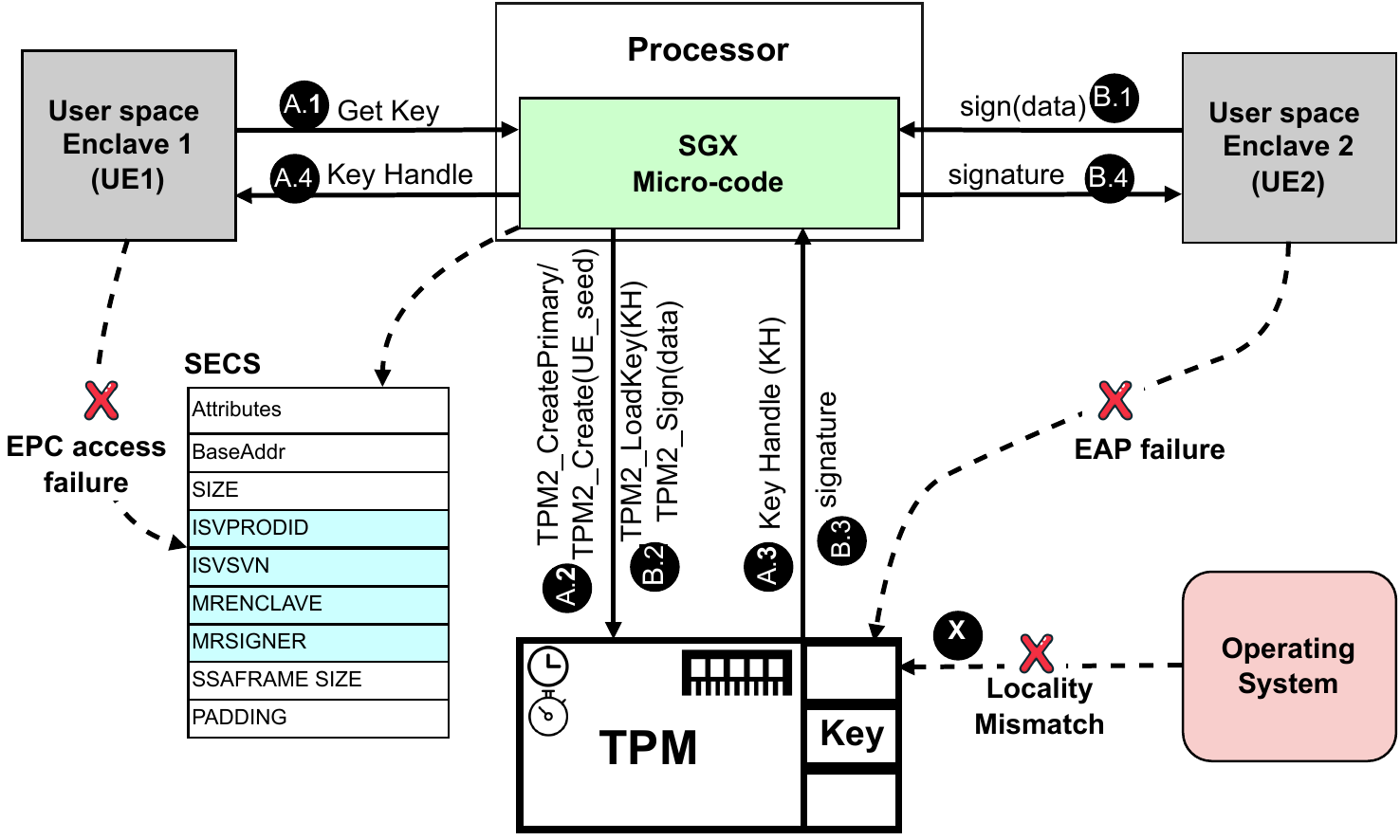} }}%
	\qquad
	\subfloat[\ournameplain \lstinline{EGETKEY} key derivation mechanism]{{\includegraphics[width=4cm]{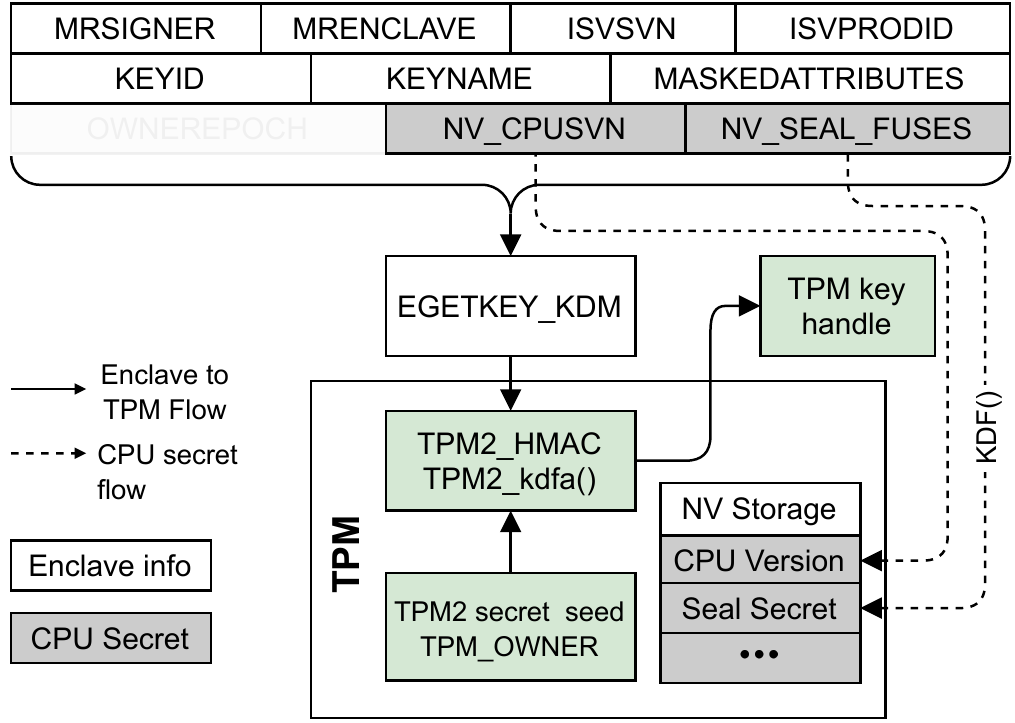} }}%
	\caption{Design and implementation of \ournameplain}%
	\label{fig:tpmconnle}%
\end{figure}

\textbf{TPM Command Mapping (\cheading{FC1}).}
Figure~\ref{fig:tpmconnle}.a shows the \ourname workflow to use the TPM as the backend for the SGX SDK functions that handle keys.
Other operations, such as reading a persistent counter from the \tpm, follow the same idea.
For persistent secure storage of the wrapped keys, an enclave can rely on the OS to store the data on the hard disk.
Creating and using counters is similar to key handling.
As TPM counters are implemented in the \tpm's NVM, creating a new counter equals creating a new dedicated NVM space with \lstinline{TPM_NVDefineSpace} and returning a handle to the enclave.
Via this handle, the enclave can read or increment the TPM-managed counter.
To retrieve the time, TPM's \lstinline{GetTime} or \lstinline{Readclock} can be used.
\tpm provides a secure clock signal with the granularity of \SI{30}{\nano\second} (LPC bus bandwidth is \SI{33}{\mega\hertz}).

For key handling, \tpm offers adequate secrets and functionalities to achieve the same bindings of keys as \sgx (\cf Figure~\ref{fig:tpmconnle}.b).
For example, \tpm's \lstinline{TPM2_OWNERSHIP} can replace the \sgx \lstinline{OWNERSHIP} or the CPU can share the \lstinline{CPUSVN} with the \tpm that can be used as KDF input (Figure~\ref{fig:tpmconnle}.b).
\tpm-generated keys can be bound to the specific \tpm{}s through \tpm secret seeds (i.e., \lstinline{TPM2_CreatePrimary} or \lstinline{TPM2_Create} for non-migratable keys).
To bind generated keys in \ourname to both CPU and \tpm, \sgx sends a secret derived from \lstinline{SEAL_FUSES} to the \tpm as input to the \tpm key generation.
Other enclave-related information 
are available in the \lstinline{SECS} created by \sgx for every enclave.
More details on the command mapping between \sgx and \tpm are available in Appendix~\ref{appendix:harpocrates}.

\textbf{Enclave Authorization (\cheading{SC2}).}
\ourname uses \tpm's extended authorization policies (EAP) to ensure that one enclave cannot have unauthorized access to another enclave's \tpm entities.
EAP policies are set during the creation of a \tpm entity, such as a key.
The CPU in \ourname dictates the EAP of newly created \tpm entities. 
It handles the policy sessions with the \tpm, supplying the necessary information for authorization from the key-derivation material.
With EAP, we can represent the same policies reflected in the key-derivation material selection in default SGX.
For example, if a key is created with \lstinline{MRSIGNER} selected but not \lstinline{MRENCLAVE}, \ie it can be derived by all enclaves of the same developer, we represent this in an EAP that requires the enclave's \lstinline{MRSIGNER} value.
When using the key, the CPU supplies the current enclave's \lstinline{MRSIGNER} value to the \tpm policy session.
Only if it matches the value set in the EAP at key creation time can that enclave use the key.

\subsection{Limitations of the \ournameplain Implementation}
\label{securityanalysis}


Our proof-of-concept implementation demonstrates that \tpm and \sgx are very amenable for integration, leading to improved enclave security (\cf \Cref{usecases}).
Our security discussion motivates further research into more secure integration of coprocessors with CPUs.
%
In our proof-of-concept implementation, the CPU uses an end-to-end encrypted channel with pre-shared keys to the TPM (\lstinline{TPM_TakeOwnership}).
Hence, we rely on a non-compromised chipset to, \eg prevent cuckoo attacks~\cite{Parno2008}.
%
A coprocessor physically integrated into the CPU, such as Microsoft Pluton~\cite{msPluton}, can remove the dependency on the chipset for a secure, authenticated connection. 
While we did not attempt such a tighter integration for the proof-of-concept in this paper, we provide functional objectives and requirements for a secure integration between a coprocessor and an enclave. 
More details are available in Appendix~\ref{appendix:securityanalysis}.


%

\section{Case Studies}  
\label{usecases}
%

In this section, we present two case studies using \ourname. 
We demonstrate how \ourname protects the enclave life cycle by storing all long-term secrets in the \tpm. 
We also show how to strengthen mitigations against microarchitectural attacks by reducing the amount of data to protect and limiting enclave restarts. 
\subsection{\ournameplain-backed Enclave Management}
\label{UC:tpmbasedenclavebuilding}

\begin{figure}[t]%
	\centering
	\subfloat[\tpm communication from user space enclaves for \sgx operations.]{{\includegraphics[width=5cm]{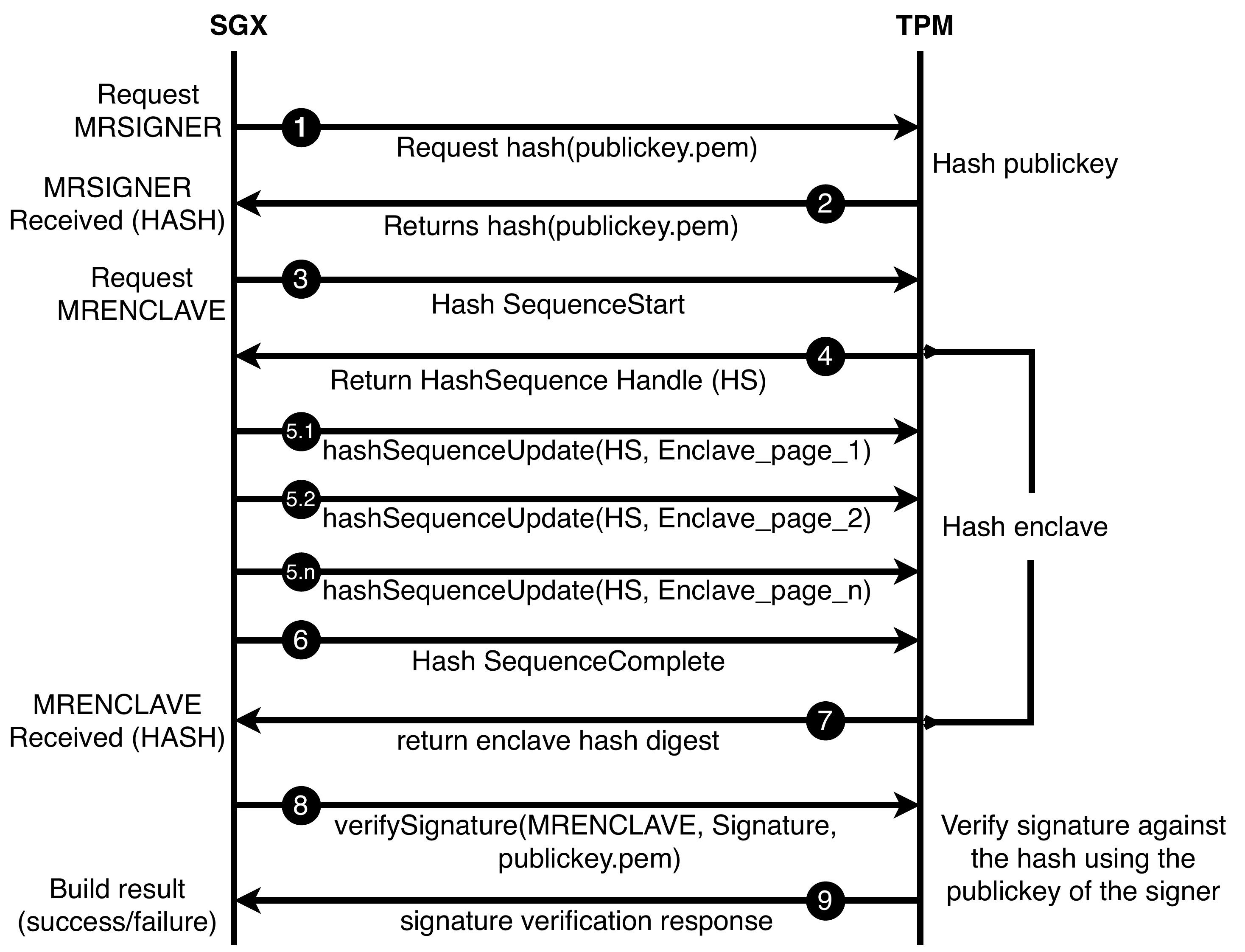} }}%
	\qquad
	\subfloat[\ournameplain \lstinline{EGETKEY} key derivation mechanism]{{\includegraphics[width=6cm]{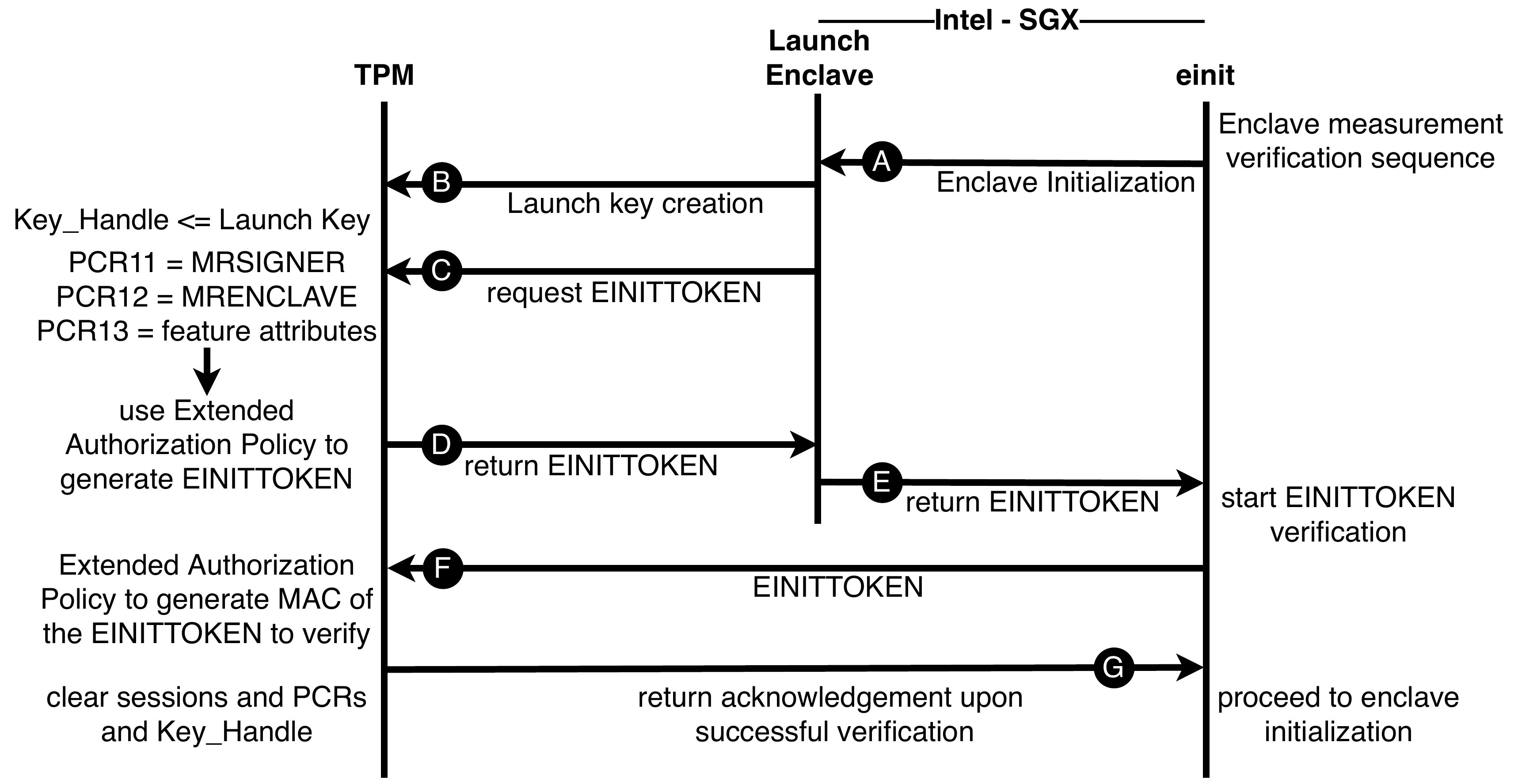} }}%
	\caption{Enclave-related use-cases for \ournameplain}%
	\label{fig:usecaseenclacveverification}%
\end{figure}

\noindent
\textbf{Enclave Creation.} 
\Cref{fig:usecaseenclacveverification}.a shows the 
two-step process of \tpm-backed enclave creation: (i)~allocating enclave pages in EPC and addition of code and data to those pages, and (ii)~measuring page contents (\lstinline{MRENCLAVE}) and verification of the measurement against a signed reference value. 
With \ourname, the \tpm creates and verifies \lstinline{MRSIGNER} and \lstinline{MRENCLAVE}.
These operations require hashing of \lstinline{MRSIGNER} using \tpm commands like \lstinline{TPM2_HashSequenceStart}, \lstinline{TPM2_HashSequenceUpdate} and lastly \lstinline{TPM2_HashSequenceComplete}.
The \tpm returns the hash of the measured enclave pages, \ie \lstinline{MRENCLAVE}.
\sgx verifies the measurement of the enclave code (using the command \lstinline{TPM2_VerifySignature}) with the reference value signed by the creator of the enclave using the creator's public key.
If the values are the same, the enclave creation is successful.


\label{UC:tpmBasedEnclaveVerification}

\noindent\textbf{Enclave Launch.}
A successfully created enclave is launched using the \lstinline{EINIT} command.
Vanilla \sgx employs a complex launch-control mechanism involving the LE, which requires a launch key (LK)~\cite{CostanD2016}. 
By default, the LK is derived using the same key derivation used for sealing keys, 
and transferred between the trusted runtime and LE via microarchitectural buffers.
Transient-execution attacks~\cite{Bulck2018,Schwarz2019} attacked these buffers to extract the launch key. 
\ourname replaces this unprotected buffer transfer by encapsulating the key inside the \tpm and releasing it upon successful authorization.
We implement the launch control using \tpm (\cf \Cref{fig:usecaseenclacveverification}.b).
The launch process starts when \lstinline{EINIT} requests an enclave initialization (\cf \Cref{fig:usecaseenclacveverification}.b) from the LE.
The LE issues an LK request to the \tpm with the \lstinline{TPM2_CreatePrimary} command.
Note that this process can also be ported to Intel DCAP.


The related enclave information from Enclave SECS is passed to the \tpm.
The \tpm creates a key using the \lstinline{EINITTOKEN} KDM as supplied by the CPU.
\sgx also resets \tpm PCRs and extends the enclave information into those PCRs (e.g., PCR 11-13).
The PCR extension 
is a well-known procedure used in, e.g., Flicker~\cite{McCune2008}, other solutions for proof-of-execution~\cite{perez2006vtpm}, and measured boot mechanisms~\cite{intelTXT2021}.
After the \tpm returns a key handle, an \lstinline{EINITTOKEN} generation request is issued, wrapped in an EAP session using the enclave identity information as policy.
Therefore, the authorization succeeds only if the correct enclave information was extended into the PCRs.
The \tpm creates the \lstinline{EINITTOKEN}, an HMAC of the enclave identity information, using the launch key loaded into the \tpm.
The \lstinline{EINITTOKEN} is returned to \lstinline{EINIT} (\circledb{D}) from the LE.
\lstinline{EINIT} receives the \lstinline{EINITTOKEN} and sends it to the \tpm for verification (\circledb{F}).
After verification, the \tpm returns an acknowledgment of success to \lstinline{EINIT} (\circledb{G}) to proceed, setting the enclave's \lstinline{INIT} attribute to true.
This enables a ring~3 application to execute the enclave's code using \sgx instructions.
The used PCRs are reset to their predefined values, which is possible because the code runs at locality~4.

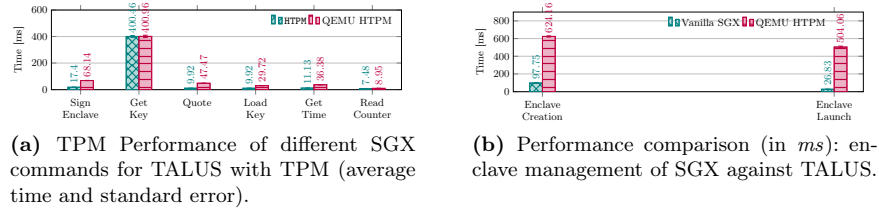
\begin{figure}[t]%
	\centering
	\resizebox{0.45\hsize}{!}{
	\subfloat[\tpm Performance of different SGX commands for \ournameplain with \tpm (average time and standard error).]{{\pgfplotstableread[
        col sep=comma,
            ]{figures/e2e.csv}\datatable

\begin{tikzpicture}[scale=0.5]
  \begin{axis}[
    compat=newest,
    ybar=0pt,
    bar width=11pt,
    xtick=data,
    legend columns=4,
    legend style={at={(0.99,0.97) }, anchor=north east,draw=none,fill=none,font=\small},
    ylabel={Time [ms]},
    ylabel style={text width=2cm,align=center},
    xticklabels from table={\datatable}{NAME},
    width=1.0\hsize,
    ymin=-1,
    ymax=600,
    ymajorgrids,
    nodes near coords,
    every node near coord/.append style={/pgf/number format/1000 sep=,rotate=90, anchor=west},
    x tick label style={rotate=0,text width=33,align=center},
    x tick label style={font=\small},
    height=4cm
  ]
  \definecolor{red2}{RGB}{191,0,64}
  
  \addplot+[teal,fill=teal!30,postaction={pattern=crosshatch,pattern color=teal},mark=none, error bars/.cd, y dir=both, y explicit] table[x expr=\coordindex,y=HTPM, y error=HTPM-SD] {\datatable};
    \addlegendentry{\texttt{HTPM}};
  \addplot+[red2,fill=red2!30,postaction={pattern=horizontal lines,pattern color=red2},mark=none, error bars/.cd, y dir=both, y explicit] table[x expr=\coordindex,y=QEMU-HTPM, y error=QEMU-HTPM-SD] {\datatable};
      \addlegendentry{QEMU HTPM};

  \end{axis}
\end{tikzpicture} }}}%
	\qquad
	\resizebox{0.45\hsize}{!}{
	\subfloat[Performance comparison (in \textit{ms}): enclave management of \sgx against \ournameplain.]{{\pgfplotstableread[
        col sep=comma,
            ]{figures/sgx.csv}\datatable

\begin{tikzpicture}[scale=0.5]
  \begin{axis}[
    compat=newest,
    ybar=0pt,
    bar width=11pt,
    xtick=data,
    legend columns=4,
    legend style={at={(0.91,0.97) }, anchor=north east,draw=none,fill=none,font=\small},
    ylabel={Time [ms]},
    ylabel style={text width=2cm,align=center},
    xticklabels from table={\datatable}{NAME},
    width=1.0\hsize,
    ymin=-1,
    ymax=900,
    ymajorgrids,
    nodes near coords,
    every node near coord/.append style={/pgf/number format/1000 sep=,rotate=90, anchor=west},
    x tick label style={rotate=0,text width=43,align=center},
    x tick label style={font=\small},
    height=4cm
  ]
  \definecolor{red2}{RGB}{191,0,64}
  
  \addplot+[teal,fill=teal!30,postaction={pattern=crosshatch,pattern color=teal},mark=none, error bars/.cd, y dir=both, y explicit] table[x expr=\coordindex,y=SGX, y error=SGX-SD] {\datatable};
    \addlegendentry{Vanilla SGX};
  \addplot+[red2,fill=red2!30,postaction={pattern=horizontal lines,pattern color=red2},mark=none, error bars/.cd, y dir=both, y explicit] table[x expr=\coordindex,y=QEMU-HTPM, y error=QEMU-HTPM-SD] {\datatable};
      \addlegendentry{QEMU HTPM};

  \end{axis}
\end{tikzpicture} }}}%
	\caption{2 \ournameplain performance evaluation}%
	\label{fig:measurements_paper}%
\end{figure}

\noindent
\textbf{Performance of Enclave Management using \ourname.}
\Cref{fig:measurements_paper}.b shows the performance of the TPM-backed functions. 
Enclave creation, which includes allocating enclave pages, measuring page contents, and verifying the measurements, takes on average \SI{624.16}{\milli\second} with \ourname and a hardware TPM (QEMU-HTPM).
Compared to vanilla SGX, which also takes \SI{97.75}{\milli\second}, this is only an overhead of \SI{526.41}{\milli\second}.
Given that the creation of an enclave is a one-time event in the life cycle of an enclave and does not affect any operation at runtime, this overhead is likely amortized over the runtime of the enclave.

\noindent
\textbf{SGX Attestation (FC2).}
\label{sgxattestation}
%
For \sgx, attestation is implemented in the QE.
\sgx employs local attestation to prove an enclave's identity to the QE.
The QE uses the attestation keys provisioned to the platform to attest the platform information and the attested enclave's \lstinline{MRENCLAVE}.
A \tpm naturally supports attestation using attestation keys, however, only of \tpm-internal data (e.g., PCR values or \tpm entities). 
With \ourname, we adapt the mechanism implemented by Intel and AMD for DRTM/Late-launch, where the platform attests with the TPM a small piece of code measured by the CPU.
DRTM uses PCR17 of the \tpm for measurement attestation.
The CPU can only reset PCR17 at locality~4.
Hence, a verifier is assured that the attested measurement in PCR17 can only come from the CPU during DRTM.
In \ourname, we designate PCR21 for \sgx attestation and set an EAP on this PCR that allows only locality~4 to read, extend, and reset this PCR.
The \tpm can attest this policy to a remote verifier to ensure them about this policy.
During \sgx attestation, the microcode resets PCR21 and extends it with the measurement of the QE (i.e., \lstinline{MRENCLAVE} of the QE) and the report generated by the QE.
A remote verifier can use the attested PCR21 value to check for a trusted QE and the proper report, \ie \lstinline{MRENCLAVE} and optionally supplied data to the report.
Note that the EPID attestation used by \sgx~\cite{intelSGXRemAtt2016} is an extended version of \tpm's DAA and can be modeled entirely using DAA~\cite{Brickell2012}.
Simply extending the enclave \lstinline{MRENCLAVE} into a PCR and attesting this PCR is insufficient without ensuring that the \lstinline{MRENCLAVE} is correct and reported by a trusted entity.


\noindent
\textbf{Performance of Other Co-processor Functions}
We evaluate the runtime of \lstinline{Sign Enclave}, \lstinline{Get Key}, \lstinline{Quote}, \lstinline{Load key}, \lstinline{Get Time} and \lstinline{Read Counter} provided by \ourname.
As a baseline, we measure the time it takes the hardware TPM (HTPM) to execute these primitives.
\Cref{fig:measurements_paper}.a shows the average execution time 
over \SI{1000}{} measurements and a \SI{95}{\percent} confidence interval.
Communication between the TPM and SGX adds a small average overhead between \SI{0.49}{\milli\second} (generating a 2048-bit RSA key) and \SI{50.77}{\milli\second} (enclave signing).

\ournameplain running with a hardware \tpm adds an average overhead of \SI{98.61}{\milli\second} $\pm$ \SI{1.95}{\milli\second}.
Note that the overall runtime overhead of an enclave depends on its workload, \ie how often these commands are executed.

\begin{figure}[t]
	\centering
	\resizebox{\hsize}{!}{
		\pgfplotstableread[
        col sep=comma,
            ]{figures/overhead.csv}\datatable

\begin{tikzpicture}[scale=0.08]
  \definecolor{red2}{RGB}{191,0,64}
  
  \begin{axis}[
    xbar stacked,
    xmin=0,
    xmax=480,
    ytick=data,
    style={font=\footnotesize},
    xlabel={Time [ms]},
    height=4.75cm,
    width=\hsize,
    bar width=7pt,
    yticklabels from table={\datatable}{NAME}
    ]
    \addplot [red,fill=red!30,postaction={pattern=crosshatch,pattern color=red}] table [x=CORRECTED, meta=NAME,y expr=\coordindex] {\datatable};
    \addplot [teal,fill=teal!30,postaction={pattern=horizontal lines,pattern color=teal},
    nodes near coords={{\scriptsize ~+\pgfmathprintnumber[precision=2]{\pgfplotspointmeta}\,ms}},
    nodes near coords align={anchor=west},
    every node near coord/.append style={
        black,
        outer sep=\pgflinewidth
    }] table [x=QEMU, meta=NAME,y expr=\coordindex] {\datatable};
    \legend{Base,QEMU}
  \end{axis}
\end{tikzpicture}
	}
	\caption{The total runtime of the commands split into base execution time and the overhead added by QEMU. }
	\label{fig:qemu-overhead}
\end{figure}
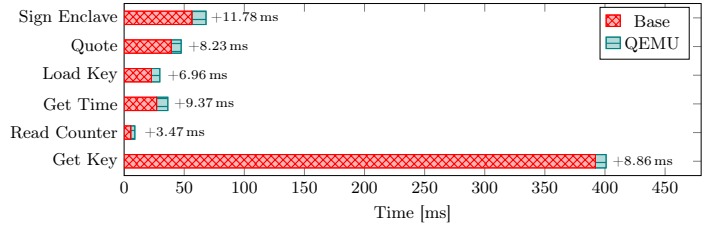

\noindent
\textbf{Data Encryption using \ournameplain.}
We evaluate a real-world use case that encrypts data using AES without leaking the key, even in the presence of transient execution attacks (\cf \Cref{UC:mitigations}).
Our application uses a 128-bit AES key securely stored in the TPM, only fetched when encrypting user-provided data.
To ensure no leakage of round keys via the SSA~\cite{Schwarz2019}, we execute the round-key derivation and encryption within a hardware transaction~\cite{Gruss2017Cloak}.
The total runtime of encrypting \SI{4}{\kilo B} of data and cleaning up any secret state is \SI{1.66}{\micro\second}$\pm$\SI{0.001}{\micro\second}, excluding fetching the key from the TPM.
The overhead from \ourname, \ie securely getting the key, is \SI{58.43}{\milli\second}$\pm$\SI{1.45}{\milli\second}.
As a baseline, we compare the runtime to a variant where the key is not fetched from the TPM but unsealed from the disk.
This (insecure) variant has an average runtime of \SI{199.21}{\micro\second}$\pm$\SI{0.45}{\micro\second}.
Note that the one-time overhead is amortized if the enclave runtime increases, \eg if larger amounts of data are encrypted.

Since only Intel can implement a native version of \ourname, and there is no cycle-accurate emulator that supports SGX, we can only provide an estimate for such a version.
\Cref{fig:qemu-overhead} shows the overhead added by QEMU for the \ourname commands, adding an overhead between \SIrange{5}{10}{\milli\second} (avg. of \SI{6.82}{\milli\second}).
This overhead constitutes between \SIrange{2.21}{38.77}{\percent} (avg. of \SI{21.60}{\percent}).
We assume that commands in a native \ourname implementation are around \SI{20}{\percent} faster.

\subsection{Impeding Microarchitectural Attacks}
\label{UC:mitigations}

SGX enclaves are a constant target of microarchitectural attacks~\cite{vanbulck2020phd,Schwarz2020How}. 
The property that enclaves can be started arbitrarily often makes it challenging to write side-channel-resilient code~\cite{Schwarz2020How}.
Furthermore, with transient-execution attacks such as Foreshadow~\cite{Bulck2018}, Spectre~\cite{Kocher2018}, RIDL~\cite{ridl2019}, ZombieLoad~\cite{Schwarz2019}, and architectural vulnerabilities such as \AE PIC~\cite{Borrello2022AEPIC}, attackers can leak sensitive data from internal CPU buffers despite side-channel-resilient code.

\textbf{Preventing Transient-Execution Attacks.}
\tpms are assumed to be resilient against other forms of microarchitectural attacks since no untrusted code can access the hardware of a \tpm.
Further, by design, \tpm does not release any secret keys managed by \tpm to the outside, but only \textit{key handles}.
However, sometimes the \tpm needs to release secret data to the enclave (e.g., a decrypted symmetric key).
With \ourname, data is loaded directly into CPU registers.
No transient-execution attack against CPU general-purpose registers has been demonstrated~\cite{Canella2019A}.
Note that Meltdown attacks were only shown against system registers~\cite{rogueSystemRegRead2018,Canella2019A} and floating-point and the upper half of SIMD registers in specific scenarios~\cite{Stecklina2018,Moghimi2020medusa,Intel2020vector}.
Hence, as long as a secret is only stored in, \eg an MPX register (\lstinline{BND0}-\lstinline{BND3}), it cannot be leaked using a transient-execution attack.
Otherwise, Meltdown mitigations, such as KPTI, would also be ineffective.

\textbf{Proof-of-concept Evaluation.}
As a proof of concept, we reproduce the AES-NI encryption from ZombieLoad~\cite{Schwarz2019}.
With \ourname, we can load the AES key from the \tpm directly into the CPU registers without requiring a memory load.
Hence, the attack vector used by Schwarz~\etal\cite{Schwarz2019} is mitigated.
To mitigate the remaining attack vector, the storing and loading of the \lstinline{XMM} registers in the SSA, we rely on Cloak~\cite{Gruss2017Cloak} to not leak any intermediate results from the registers to memory.
We verify that the plain AES key is never stored in memory by inspecting the memory.
Further, we are certain that the key is not stored in any vulnerable microarchitectural element used for interacting with the memory, such as the store buffer or line-fill buffer, preventing leakage via transient-execution attacks. 
However, we cannot exclude the existence of unknown buffers that are on the data path in Cloak~\cite{Gruss2017Cloak} and that might become vulnerable in the future. 

\textbf{Limiting Precise Execution Control \& Strengthening Countermeasures.}
Due to the strong attacker model, \sgx enclaves can be interrupted at an arbitrary point, allowing precise execution control~\cite{Schwarz2020How}.
With SGX-step~\cite{sgxstep2017}, enclaves can be interrupted after every instruction, allowing to amplify side-channel leakage.
Constant interruptions result in constantly storing and loading of the enclave state, resulting in more reliable transient-execution attacks~\cite{Bulck2018,Schwarz2019}.
By design, \ourname does not store secrets stored in the MPX registers in plain memory, preventing leakage of these values (\cf \Cref{harpocrates}).
While \ourname cannot directly prevent precise execution control, its persistent storage can track how often an enclave was interrupted.
Although enclaves can detect interrupts 
via overwritten values in the SSA~\cite{Oleksenko2018varys,Chen2019defeating}, they cannot store this information across enclave restarts.
With \ourname, an enclave can track the number of interrupts across enclave restarts.
Due to this persistent storage, an enclave can refuse to start if it suffers from an excessive number of interrupts.

Generally, \ourname allows enclaves to keep information across restarts, strengthening state-of-the-art countermeasures against microarchitectural attacks.
T-SGX~\cite{Shih2017tsgx}, Varys~\cite{Oleksenko2018varys}, or D\'ej\`a Vu~\cite{Chen2017detecting} drastically reduce the observable leakage during one enclave run.
However, since they cannot prevent arbitrary enclave restarts, leakage is still possible~\cite{Jiang2020monitoring}.
Using secure counters of \ourname strengthens such countermeasures to prevent an enclave from starting if too many abnormal events have been observed during execution.

\textbf{Proof-of-concept Evaluation.}
We implement the restart limitation in the sample enclave of T-SGX~\cite{Shih2017tsgx}.
The enclave first increments a counter stored in the \tpm and retrieves the current value.
This value is the number of times the enclave has been started.
Only if the current counter value is below an enclave-defined threshold the enclave continues to provision the secrets.
The limit can be obtained from a remote server to increase the number of allowed executions over time gradually.
Contrary to the number of enclave executions, storing this threshold in a sealed data blob is possible.
A rollback attack would only decrease the number of remaining enclave executions, providing no advantages to an attacker. 
As the check only happens once at enclave startup, this is a one-time overhead. 
With T-SGX, the time it takes to create and launch the enclave is \SI{19.66}{\milli\second}$\pm$\SI{0.016}{\milli\second} ($n$ = 1000). 
Increasing, reading, and comparing the timer with \ourname takes on average \SI{17.45}{\milli\second}$\pm$\SI{0.23}{\milli \second}.


%


\section{Other Platforms}
\label{otherplatfrom}
%

\ourname shows how a co-processor can be integrated with a TEE on x86. 
%
%
Other platforms, such as ARM and RISC-V, can also benefit from our requirement analysis. 
For example, ARM TrustZone supports co-processors such as Google Titan or Apple T2 but with limited use cases such as disk encryption, key generation or encryption. 
On RISC-V, Keystone Enclaves and RoCC (Rocket chip coprocessor) are available on the Boom core~\cite{CelioEECS2017} and Rocket core~\cite{AsanovicEECS2016}.
Hence, also on RISC-V, integrating the co-processor with enclaves can provide better security guarantees. 
A detailed discussion on how other platform can benefit from a \ournameplain implementation is available in Appendix~\ref{app:otherplatfrom}.

\section{Conclusion}
\label{conclusion}
%

We showed that secure enclaves, such as \sgx, can benefit from secure coprocessors, such as a \tpm, if they are securely integrated.
With \ourname, we presented a design that supports secure side-channel-resilient communication between TEEs and cryptographic coprocessors.
We presented a proof-of-concept implementation based on a hardware \tpm and \sgx, demonstrating how a \tpm can protect the \sgx infrastructure credentials during enclave building and launching, and how such a design impedes microarchitectural attacks on \sgx.
From our prototype, we derive crucial requirements for secure integration between TEEs and coprocessors.
We believe that the identified and solved challenges leading to our design of \ourname are valuable for future systems, such as integrating Microsoft's Pluton with enclaves, and can be transferred to other combinations of enclave technology and coprocessors, such as AMD PSP or ARM TrustZone.

%


%

%
%
%
%
%
%
%
%

\bibliographystyle{splncs04}
\bibliography{acmart}

\begin{thebibliography}{10}
\providecommand{\url}[1]{\texttt{#1}}
\providecommand{\urlprefix}{URL }
\providecommand{\doi}[1]{https://doi.org/#1}

\bibitem{AMDPSPCCC2019}
{AMD}: Uncover, understand, own - regaining control over your amd cpu (2019)

\bibitem{secureEnclaveAPPLE}
{Apple Inc.}: Apple platform security - secure enclave.
  \url{https://support.apple.com/guide/security/secure-enclave-sec59b0b31ff/web}
  (2021), accessed: 12/08/2021

\bibitem{t2APPLE}
{Apple Inc.}: Apple t2 security chip - security overview.
  \url{https://www.apple.com/mideast/mac/docs/Apple_T2_Security_Chip_Overview.pdf}
  (2021), accessed: 12/08/2021

\bibitem{armTZ}
{Arm Limited}: The trustzone hardware architecture.
  \url{https://developer.arm.com/documentation/100935/0100/The-TrustZone-hardware-architecture-?lang=en}
  (2021), accessed: 01.05.2021

\bibitem{AsanovicEECS2016}
Asanović, K., Avizienis, R., Bachrach, J., Beamer, S., Biancolin, D., Celio,
  C., Cook, H., Dabbelt, D., Hauser, J., Izraelevitz, A., Karandikar, S.,
  Keller, B., Kim, D., Koenig, J., Lee, Y., Love, E., Maas, M., Magyar, A.,
  Mao, H., Moreto, M., Ou, A., Patterson, D.A., Richards, B., Schmidt, C.,
  Twigg, S., Vo, H., Waterman, A.: The rocket chip generator. Tech. rep., EECS
  Department, University of California, Berkeley (Apr 2016),
  \url{http://www2.eecs.berkeley.edu/Pubs/TechRpts/2016/EECS-2016-17.html}

\bibitem{tpmgenie}
Boone, J.: Tpm genie: Attacking the hardware root of trust for less than \$50
  (2018), accessed: 02/13/2019

\bibitem{Borrello2022AEPIC}
Borrello, P., Kogler, A., Schwarzl, M., Lipp, M., Gruss, D., Schwarz, M.:
  {ÆPIC Leak}: Architecturally leaking uninitialized data from the
  microarchitecture. In: 31st USENIX Security Symposium (USENIX Security 22)
  (2022)

\bibitem{Bourgeat2019}
Bourgeat, T., Lebedev, I., Wright, A., Zhang, S., Arvind, Devadas, S.: Mi6:
  Secure enclaves in a speculative out-of-order processor. In: MICRO '52.
  Association for Computing Machinery (2019)

\bibitem{Brickell2012}
{Brickell}, E., {Li}, J.: Enhanced privacy id: A direct anonymous attestation
  scheme with enhanced revocation capabilities. IEEE Transactions on Dependable
  and Secure Computing  (2012)

\bibitem{Bulck2018}
Bulck, J.V., Minkin, M., Weisse, O., Genkin, D., Kasikci, B., Piessens, F.,
  Silberstein, M., Wenisch, T.F., Yarom, Y., Strackx, R.: Foreshadow:
  Extracting the keys to the intel {SGX} kingdom with transient out-of-order
  execution. In: 27th {USENIX} Security Symposium ({USENIX} Security 18) (2018)

\bibitem{fallout2019}
Canella, C., Genkin, D., Giner, L., Gruss, D., Lipp, M., Minkin, M., Moghimi,
  D., Piessens, F., Schwarz, M., Sunar, B., {Van Bulck}, J., Yarom, Y.:
  Fallout: Leaking data on meltdown-resistant cpus. In: Proceedings of the ACM
  SIGSAC Conference on Computer and Communications Security ({CCS}) (2019)

\bibitem{Canella2019A}
Canella, C., Van~Bulck, J., Schwarz, M., Lipp, M., von Berg, B., Ortner, P.,
  Piessens, F., Evtyushkin, D., Gruss, D.: {A Systematic Evaluation of
  Transient Execution Attacks and Defenses}. In: USENIX Security Symposium
  (2019), extended classification tree and PoCs at https://transient.fail/.

\bibitem{CelioEECS2017}
Celio, C., Chiu, P.F., Nikolic, B., Patterson, D.A., Asanović, K.: Boom v2: an
  open-source out-of-order risc-v core. Tech. rep., EECS Department, University
  of California, Berkeley (Sep 2017),
  \url{http://www2.eecs.berkeley.edu/Pubs/TechRpts/2017/EECS-2017-157.html}

\bibitem{Chen2019defeating}
Chen, G., Li, M., Zhang, F., Zhang, Y.: {Defeating Speculative-Execution
  Attacks on SGX with HyperRace}. In: Dependable and Secure Computing (DSC)
  (2019)

\bibitem{Chen2017detecting}
Chen, S., Zhang, X., Reiter, M.K., Zhang, Y.: Detecting privileged side-channel
  attacks in shielded execution with d{\'e}j{\'a} vu. In: AsiaCCS (2017)

\bibitem{Cloosters2020teerex}
Cloosters, T., Rodler, M., Davi, L.: Teerex: Discovery and exploitation of
  memory corruption vulnerabilities in $\{$SGX$\}$ enclaves. In: USENIX
  Security Symposium (2020)

\bibitem{Cole2020simplex}
Cole, M., Prakash, A.: Simplex: Repurposing intel memory protection extensions
  for information hiding. arXiv:2009.06490  (2020)

\bibitem{Schmidt2015}
Colin~Schmidt, A.I.: A fast parameterized sha3 accelerator.
  \url{https://people.eecs.berkeley.edu/~kubitron/courses/cs262a-F13/projects/reports/project11_report.pdf}
  (2015)

\bibitem{CostanD2016}
Costan, V., Devadas, S.: Intel sgx explained. IACR Cryptology ePrint Archive
  (2016)

\bibitem{Futral2013}
Futral, W., Greene, J.: Intel Trusted Execution Technology for Server
  Platforms: A Guide to More Secure Datacenters. Apress, Berkely, CA, USA, 1st
  edn. (2013)

\bibitem{Garmany2013prime}
Garmany, B., M{\"u}ller, T.: {PRIME: Private RSA Infrastructure for Memory-less
  Encryption}. In: ACSAC (2013)

\bibitem{titanM}
Google: Titan m makes pixel 3 our most secure phone yet.
  \url{https://www.blog.google/products/pixel/titan-m-makes-pixel-3-our-most-secure-phone-yet}
  (2018), accessed: 01/06/2020

\bibitem{chromiumTPM}
Google: Tpm usage - the chromium project.
  \url{https://www.chromium.org/developers/design-documents/tpm-usage} (2019)

\bibitem{titanM21}
{Google}: Hardware security module.
  \url{https://developer.android.com/training/articles/keystore\#HardwareSecurityModule}
  (2021), accessed: 15/12/2021

\bibitem{AndroidKeystore}
{Google}: Work with keystore entries.
  \url{https://developer.android.com/training/articles/keystore\#UserAuthentication}
  (2021), accessed: 15/12/2021

\bibitem{AndroidKeystoreSecurityFeature}
{Google}: Work with keystore entries.
  \url{https://developer.android.com/training/articles/keystore\#UserAuthentication}
  (2021), accessed: 15/12/2021

\bibitem{intelSGX2016}
{Greene, James}: Intel® smi transfer monitor (stm) user guide.
  \url{https://firmware.intel.com/content/smi-transfer-monitor-stm} (2016)

\bibitem{Gruss2017Cloak}
Gruss, D., Schuster, F., Ohrimenko, O., Haller, I., Lettner, J., Costa, M.:
  {Strong and Efficient Cache Side-Channel Protection using Hardware
  Transactional Memory}. In: {USENIX Security Symposium} (2017)

\bibitem{Han2018}
Han, S., Shin, W., Park, J.H., Kim, H.: A bad dream: Subverting trusted
  platform module while you are sleeping. In: 27th {USENIX} Security Symposium
  ({USENIX} Security 18). {USENIX} Association (2018),
  \url{https://www.usenix.org/conference/usenixsecurity18/presentation/han}

\bibitem{rogueSystemRegRead2018}
Intel: Rogue system register read (2018),
  \url{https://software.intel.com/content/www/us/en/develop/articles/software-security-guidance/advisory-guidance/rogue-system-register-read.html}

\bibitem{Intel2020vector}
Intel: {Vector Register Sampling / CVE-2020-0548 / INTEL-SA-OO329} (2020),
  \url{https://software.intel.com/content/www/us/en/develop/articles/software-security-guidance/advisory-guidance/vector-register-sampling.html}

\bibitem{LPCbusSpec}
{Intel Corporation}: {Intel Low Pin Count Interface Specification, August
  2002}.
  \url{https://www.intel.com/content/www/us/en/design/technologies-and-topics/low-pin-count-interface-specification.html}
  (2002), last accessed: 31/10/19

\bibitem{intelPTT2014}
{Intel Corporation}: Strengthening security with intel®platform trust
  technology.
  \url{https://www.intel.com/content/dam/www/public/us/en/documents/white-papers/enterprise-security-platform-trust-technology-white-paper.pdf}
  (2014)

\bibitem{intelSGXDevMan2016}
{Intel Corporation}: Intel® 64 and ia-32 architectures software developer’s
  manual.
  \url{https://www.intel.com/content/dam/www/public/us/en/documents/manuals/64-ia-32-architectures-software-developer-instruction-set-reference-manual-325383.pdf}
  (2016)

\bibitem{intelSGX}
{Intel Corporation}: Intel® software guard extensions (intel sgx).
  \url{https://software.intel.com/en-us/sgx} (2016), accessed: 15.07.2019

\bibitem{IntelSTM2019}
{Intel Corporation}: Smi transfer mode (stm).
  \url{https://software.intel.com/content/www/us/en/develop/articles/smi-transfer-monitor-stm.html}
  (2019), accessed: 06/05/2020

\bibitem{intelCSME2020}
{Intel Corporation}: Intel converged security and management engine (intel
  csme).
  \url{https://www.intel.com/content/dam/www/public/us/en/security-advisory/documents/intel-csme-security-white-paper.pdf}
  (2020)

\bibitem{csmeATTACK2018}
{Intel Corporation}: Intel-sa-00219 sgx sw developer guidance.
  \url{https://www.intel.com/content/dam/www/public/us/en/security-advisory/documents/the-intel-csme-dam-vulnerability-cve-2018-3659-and-cve-2018-3643-whitepaper.pdf}
  (2020), accessed: 11/12/2021

\bibitem{csmeATTACK2019}
{Intel Corporation}: The intel® converged security and management engine iommu
  hardware issue – cve-2019-0090 and cve-2020-0566.
  \url{https://www.intel.com/content/dam/www/public/us/en/security-advisory/documents/cve-2019-0090-whitepaper.pdf}
  (2020), accessed: 11/12/2021

\bibitem{intelTXT2021}
{Intel Corporation}: Intel® trusted execution technology (intel® txt)
  software development guide measured launch environment developer’s guide.
  \url{http://www.intel.com/content/www/us/en/software-developers/intel-txt-software-development-guide.html}
  (2021)

\bibitem{intelCounter}
{Intel Corporation}: Unable to find alternatives to monotonic counter
  application programming interfaces (apis) in intel® software guard
  extensions (intel® sgx) for linux* to prevent sealing rollback attacks.
  \url{https://www.intel.com/content/www/us/en/support/articles/000057968/software/intel-security-products.html}
  (2021), accessed: 10.01.2022

\bibitem{intelXuCode}
{Intel Corporation}: Xucode: An innovative technology for implementing complex
  instruction flows.
  \url{https://software.intel.com/content/www/us/en/develop/articles/software-security-guidance/secure-coding/xucode-implementing-complex-instruction-flows.html}
  (2021)

\bibitem{kvmSGX}
{Intel Corporation}: Kvm sgx. \url{https://github.com/intel/kvm-sgx} (2022),
  accessed: 01/06/2020

\bibitem{qemuSGX}
{Intel Corporation}: Qemu sgx. \url{https://github.com/intel/qemu-sgx} (2022),
  accessed: 01/06/2020

\bibitem{Jiang2020monitoring}
Jiang, J., Soriente, C., Karame, G.: Monitoring performance metrics is not
  enough to detect side-channel attacks on intel sgx. arXiv:2011.14599  (2020)

\bibitem{Karandikar2021}
Karandikar, S., Leary, C., Kennelly, C., Zhao, J., Parimi, D., Nikolic, B.,
  Asanovic, K., Ranganathan, P.: A hardware accelerator for protocol buffers.
  In: MICRO-54: 54th Annual IEEE/ACM International Symposium on
  Microarchitecture. MICRO '21 (2021)

\bibitem{conf/uss/Kauer07}
Kauer, B.: Oslo: Improving the security of trusted computing. In: Proc.\ 16th
  USENIX Security Symposium (SEC '07). USENIX Association (2007)

\bibitem{Kenjar2020v0ltpwn}
Kenjar, Z., Frassetto, T., Gens, D., Franz, M., Sadeghi, A.: {V0LTpwn:
  Attacking x86 Processor Integrity from Software}. In: USENIX Security
  Symposium (2020)

\bibitem{Kocher2018}
{Kocher}, P., {Horn}, J., {Fogh}, A., {Genkin}, D., {Gruss}, D., {Haas}, W.,
  {Hamburg}, M., {Lipp}, M., {Mangard}, S., {Prescher}, T., {Schwarz}, M.,
  {Yarom}, Y.: Spectre attacks: Exploiting speculative execution. In: 2019 IEEE
  Symposium on Security and Privacy (SP) (2019)

\bibitem{TPMHWattack}
Lawson, N.: Tpm hardware attacks.
  \url{https://rdist.root.org/2007/07/16/tpm-hardware-attacks/} (Jul 2007),
  accessed: 06.08.2018

\bibitem{membuster2020}
Lee, D., Jung, D., Fang, I.T., Tsai, C.C., Popa, R.A.: An off-chip attack on
  hardware enclaves via the memory bus. In: 29th {USENIX} Security Symposium
  ({USENIX} Security 20) (2020),
  \url{https://www.usenix.org/conference/usenixsecurity20/presentation/lee-dayeol}

\bibitem{keystone2020}
Lee, D., Kohlbrenner, D., Shinde, S., Asanovi\'{c}, K., Song, D.: Keystone: An
  open framework for architecting trusted execution environments. In:
  Proceedings of the Fifteenth European Conference on Computer Systems (2020),
  \url{https://dl.acm.org/doi/abs/10.1145/3342195.3387532}

\bibitem{Lee2017SGXROP}
Lee, J., Jang, J., Jang, Y., Kwak, N., Choi, Y., Choi, C., Kim, T., Peinado,
  M., Kang, B.B.: {Hacking in Darkness: Return-oriented Programming against
  Secure Enclaves}. In: USENIX Security Symposium (2017)

\bibitem{Li2021cipherleaks}
Li, M., Zhang, Y., Wang, H., Li, K., Cheng, Y.: {CIPHERLEAKS: Breaking
  Constant-time Cryptography on AMD SEV via the Ciphertext Side Channel}. In:
  USENIX Security Symposium (2021)

\bibitem{intelVirtSGX}
M., J.: Virtualizing intel® software guard extensions with kvm and qemu.
  \url{https://software.intel.com/en-us/articles/virtualizing-intel-software-guard-extensions-with-kvm-and-qemu}
  (2019), accessed: 01/06/2020

\bibitem{MateticAKDSGJC17}
Matetic, S., Ahmed, M., Kostiainen, K., Dhar, A., Sommer, D., Gervais, A.,
  Juels, A., Capkun, S.: {ROTE}: Rollback protection for trusted execution. In:
  Proc.\ 26th USENIX Security Symposium (SEC' 17). usenix (2017)

\bibitem{McCune2008}
McCune, J.M., Parno, B.J., Perrig, A., Reiter, M.K., Isozaki, H.: Flicker: An
  execution infrastructure for tcb minimization. SIGOPS Oper. Syst. Rev.
  (2008)

\bibitem{mckeen2013}
McKeen, F., Alexandrovich, I., Berenzon, A., Rozas, C.V., Shafi, H.,
  Shanbhogue, V., Savagaonkar, U.R.: Innovative instructions and software model
  for isolated execution. Hasp@ isca  \textbf{10}(1) (2013)

\bibitem{Schwarz2020How}
Michael~Schwarz, D.G.: {How Trusted Execution Environments Fuel Research on
  Microarchitectural Attacks}. In: Security \& Privacy (2020)

\bibitem{Schwarz2019Practical}
Michael~Schwarz, Samuel~Weiser, D.G.: {Practical Enclave Malware with Intel
  SGX}. In: DIMVA (2019)

\bibitem{bitlockerTPM}
{Microsoft Corporation}: Bitlocker.
  \url{https://docs.microsoft.com/en-us/windows/security/information-protection/bitlocker/bitlocker-overview}
  (2018)

\bibitem{msPluton}
{Microsoft Corporation}: Meet the microsoft pluton processor – the security
  chip designed for the future of windows pcs.
  \url{https://www.microsoft.com/security/blog/2020/11/17/meet-the-microsoft-pluton-processor-the-security-chip-designed-for-the-future-of-windows-pcs/}
  (2021)

\bibitem{Moghimi2020medusa}
Moghimi, D., Lipp, M., Sunar, B., Schwarz, M.: {Medusa: Microarchitectural Data
  Leakage via Automated Attack Synthesis}. In: {USENIX Security Symposium}
  (2020)

\bibitem{244048}
Moghimi, D., Sunar, B., Eisenbarth, T., Heninger, N.: {TPM-FAIL: {TPM} meets
  Timing and Lattice Attacks}. In: 29th {USENIX} Security Symposium ({USENIX}
  Security 20). {USENIX} Association (2020)

\bibitem{Muller2011tresor}
M{\"u}ller, T., Freiling, F.C., Dewald, A.: Tresor runs encryption securely
  outside ram. In: USENIX Security Symposium (2011)

\bibitem{Murdock2019plundervolt}
Murdock, K., Oswald, D., Garcia, F.D., Van~Bulck, J., Gruss, D., Piessens, F.:
  {Plundervolt: Software-based Fault Injection Attacks against Intel SGX}. In:
  {S\&P} (2020)

\bibitem{Oleksenko2018varys}
Oleksenko, O., Trach, B., Krahn, R., Silberstein, M., Fetzer, C.: {Varys:
  Protecting SGX Enclaves from Practical Side-channel Attacks}. In: USENIX ATC
  (2018)

\bibitem{Parno2008}
Parno, B.: Bootstrapping trust in a "trusted" platform. In: Proceedings of the
  3rd Conference on Hot Topics in Security. HOTSEC'08 (2008)

\bibitem{perez2006vtpm}
Perez, R., Sailer, R., van Doorn, L., et~al.: vtpm: virtualizing the trusted
  platform module. In: Proc. 15th Conf. on USENIX Security Symposium. pp.
  305--320 (2006)

\bibitem{Phoronix2020MPX}
Phoronix: {Intel MPX Support Is Dead With Linux 5.6} (2020),
  \url{https://www.phoronix.com/scan.php?page=news_item&px=Intel-MPX-Is-Dead}

\bibitem{Pengfei2019}
Qiu, P., Wang, D., Lyu, Y., Qu, G.: Voltjockey: Breaching trustzone by
  software-controlled voltage manipulation over multi-core frequencies. In:
  Proceedings of the 2019 ACM SIGSAC Conference on Computer and Communications
  Security. CCS '19, Association for Computing Machinery (2019)

\bibitem{ragab_crosstalk_2021}
Ragab, H., Milburn, A., Razavi, K., Bos, H., Giuffrida, C.: {CrossTalk}:
  {Speculative} {Data} {Leaks} {Across} {Cores} {Are} {Real}. In: S\&{P} (2021)

\bibitem{ridl2019}
van Schaik, S., Milburn, A., Österlund, S., Frigo, P., Maisuradze, G., Razavi,
  K., Bos, H., Giuffrida, C.: {RIDL}: Rogue in-flight data load. In: S\&{P}
  (May 2019)

\bibitem{Schwarz2019}
Schwarz, M., Lipp, M., Moghimi, D., Van~Bulck, J., Stecklina, J., Prescher, T.,
  Gruss, D.: {ZombieLoad}: Cross-privilege-boundary data sampling. In: CCS
  (2019)

\bibitem{Shih2017tsgx}
Shih, M.W., Lee, S., Kim, T., Peinado, M.: {T-SGX: Eradicating
  controlled-channel attacks against enclave programs}. In: NDSS (2017)

\bibitem{intelSGXRemAtt2016}
{Simon Johnson and Vinnie Scarlata and Carlos Rozas and Ernie Brickell and
  Frank Mckeen}: Intel software guard extensions: Epid provisioning and
  attestation services.
  \url{https://software.intel.com/sites/default/files/managed/57/0e/ww10-2016-sgx-provisioning-and-attestation-final.pdf}
  (2016)

\bibitem{Sparks2008}
{Sparks, Sherri, Embleton, Shawn, Zou, Cliff}: Smm rootkits: A new breed of os
  independent malware.
  \url{http://www.eecs.ucf.edu/~czou/research/SMM-Rootkits-Securecom08.pdf}
  (2008)

\bibitem{Stecklina2018}
Stecklina, J., Prescher, T.: {LazyFP: Leaking FPU Register State using
  Microarchitectural Side-Channels}. arXiv:1806.07480  (2018)

\bibitem{AzureSphere}
Stiles, D.: The hardware security behind azure sphere. IEEE Micro  (2019)

\bibitem{TPM2Spec}
{Trusted Computing Group}: Tpm 2.0 library specification.
  \url{https://trustedcomputinggroup.org/resource/tpm-library-specification/}
  (2016)

\bibitem{TPM2Arch}
{Trusted Computing Group}: Trusted platform module library part 1:
  Architecture.
  \url{https://trustedcomputinggroup.org/wp-content/uploads/TCG_TPM2_r1p59_Part1_Architecture_pub.pdf}
  (2019)

\bibitem{vanbulck2020phd}
Van~Bulck, J.: Microarchitectural Side-Channel Attacks for Privileged Software
  Adversaries. Ph.D. thesis, KU Leuven (2020)

\bibitem{VanBulck2020LVI}
Van~Bulck, J., Moghimi, D., Schwarz, M., Lipp, M., Minkin, M., Genkin, D.,
  Yarom, Y., Sunar, B., Gruss, D., Piessens, F.: {LVI: Hijacking Transient
  Execution through Microarchitectural Load Value Injection}. In: S\&P (2020)

\bibitem{VanBulck2019tale}
Van~Bulck, J., Oswald, D., Marin, E., Aldoseri, A., Garcia, F.D., Piessens, F.:
  A tale of two worlds: Assessing the vulnerability of enclave shielding
  runtimes. In: CCS (2019)

\bibitem{sgxstep2017}
Van~Bulck, J., Piessens, F., Strackx, R.: Sgx-step: A practical attack
  framework for precise enclave execution control. In: Proceedings of the 2nd
  Workshop on System Software for Trusted Execution. SysTEX'17 (2017)

\bibitem{Wampler2019exspectre}
Wampler, J., Martiny, I., Wustrow, E.: Exspectre: Hiding malware in speculative
  execution. In: NDSS (2019)

\bibitem{Weichbrodt2016}
Weichbrodt, N., Kurmus, A., Pietzuch, P., Kapitza, R.: {AsyncShock: Exploiting
  Synchronisation Bugs in Intel SGX Enclaves}. In: ESORICS (2016)

\bibitem{Wojtczuk2009}
{Wojtczuk, Rafal, Rutkowska, Joanna}: Attacking intel® trusted execution
  technology.
  \url{https://invisiblethingslab.com/resources/bh09dc/Attacking\%20Intel\%20TXT\%20-\%20paper.pdf}
  (2009)

\bibitem{controlledsidechannel}
{Xu}, Y., {Cui}, W., {Peinado}, M.: Controlled-channel attacks: Deterministic
  side channels for untrusted operating systems. In: ssp15. IEEE Computer
  Society (2015)

\bibitem{Zhang2016}
Zhang, F., Zhang, H.: Sok: A study of using hardware-assisted isolated
  execution environments for security. In: Proceedings of the Hardware and
  Architectural Support for Security and Privacy 2016. HASP 2016, Association
  for Computing Machinery (2016)

\bibitem{Zhang2020exploring}
Zhang, T., Koltermann, K., Evtyushkin, D.: Exploring branch predictors for
  constructing transient execution trojans. In: ASPLOS (2020)

\end{thebibliography}

\begin{subappendices}
	\renewcommand{\thesection}{\Alph{section}}%
	\newpage

	\begin{figure}[t]
		\centering
		\includegraphics[width=0.47\textwidth]{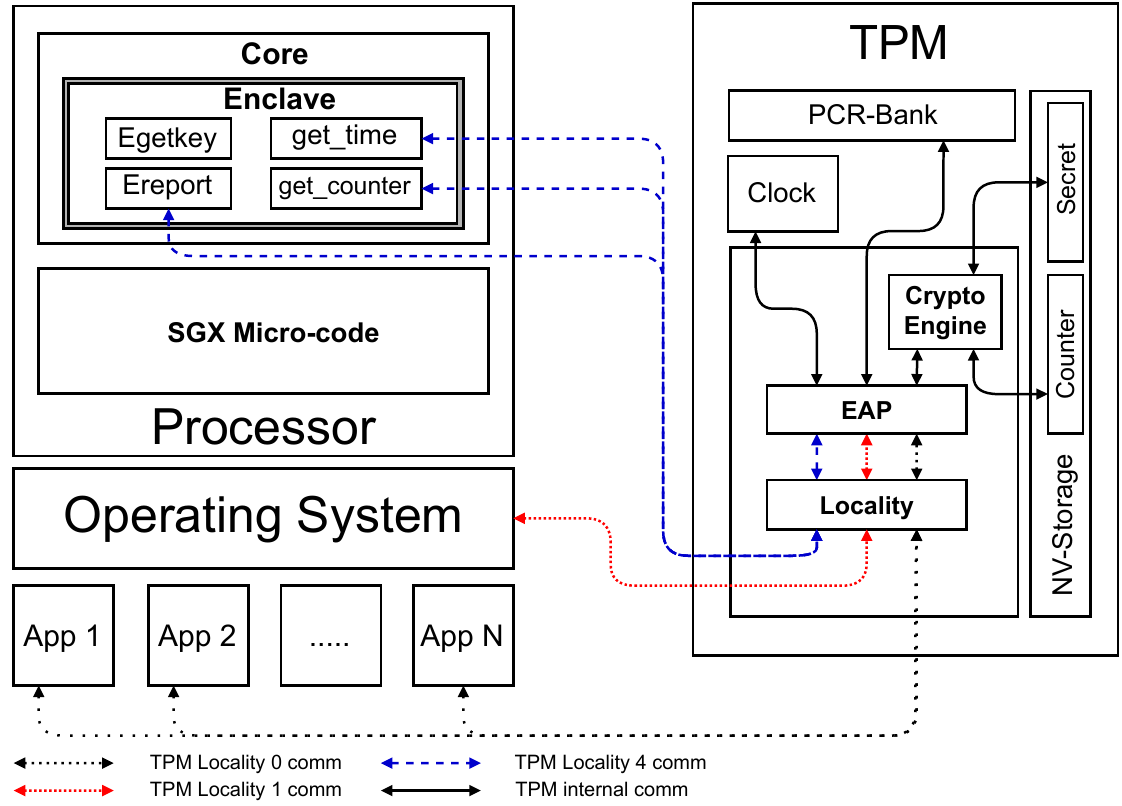} 
		\caption{High-level idea for integrating SGX enclaves with TPM as coprocessor. Access control to the \tpm entities are controlled using Locality for different levels (ring) of software and using EAP for different software working in the same level (ring); EAP = Extended Authorization Policies.}
		\label{fig:highlevelfigure}
	\end{figure}
	
	\begin{table}[t]
		\centering
		\begin{tabular}{ll}
			\toprule
			\textbf{\sgx command/operation} & \textbf{\tpm commands}   \\
			\midrule
			\small{Sgx\_get\_key} & \small{TPM2\_CreatePrimary/TPM2\_Create} \\
			\small{Sgx\_get\_trusted\_time} & \small{TPM2\_GetTime/TPM2\_Readclock} \\
			\small{Sgx\_create\_monotonic\_counter} & \small{TPM2\_NVDefineSpace} \\
			\small{Encryption} & \small{(TPM2\_Load,) TPM2\_EncryptDecrypt} \\
			\small{Sign} & \small{(TPM2\_Load,) TPM2\_Sign} \\
			\small{Quote} & \small{(TPM2\_Load,) TPM2\_Quote} \\
			\bottomrule \\
		\end{tabular}
		\caption{\sgx --\tpm commands mapping}
		\label{table:commMap}
	\end{table}

	\section{\ournameplain Implementation}
	\label{app:systemDesign}
	%

In this section, we briefly introduce implementation details of a proof-of-concept of \ourname based on \sgx and a hardware \tpm. An in-depth discussion is available in \Cref{app:systemDesign}.   
We first show the functionality and all security guarantees using the Intel SGX emulator~\cite{intelVirtSGX} and a hardware \tpm, as this allows implementing the entire design of \ourname. 
For the performance evaluation, we instead use a hardware SGX enclave in combination with the same hardware \tpm, with the limitation that the communication channel is not protected against a malicious OS. 
The reason for the two parts is that only Intel can update microcode.
Hence, we cannot implement all changes required for the secure communication channel in off-the-shelf hardware. 

%

\subsection{Connecting SGX and TPM}
\label{appendix:harpocrates}

The connection between \sgx and \tpm has to fulfill the challenges outlined in \Cref{challengesforintegration}.
In this section, we present the implementation details of the communication channel, which is illustrated in \Cref{fig:highlevelfigure}.

\noindent
\linebreak
\textbf{Channel between \sgx and \tpm (\cheading{SC1}).} \label{channelBetweenSGXandTPM}
Typically, the OS provides the \tpm as an MMIO device to system and user-space software. 
However, \ourname \textit{cannot} rely on the untrusted OS to communicate between the \tpm and enclaves. 
For our proof-of-concept implementation, we rely on the end-to-end-encrypted programmed I/O channel between the CPU and the hardware \tpm. 
This channel is controlled by \txt using an access control mechanism called \textit{Locality} offered by the \tpm through \textit{\tpm Locality Address Mapping}~\cite{intelTXT2021}. 
\tpm localities indicate the source of the command within the platform. 
Locality 0 is full public access, locality 1 is the OS. 
Higher localities (up to locality 4) correspond to the highest privilege levels, \ie hardware and microcode, including \sgx. 
In \ourname, localities ensure the vertical access control to the \tpm (\eg software, OS), while command authorization (\cf \Cref{sec:integration}) ensures the horizontal access control (\ie different enclaves).

The programmed I/O communication channel allows the CPU to choose where to store (sensitive) data received from the \tpm, \eg directly into CPU registers (\cheading{SC3}). 
In contrast, for direct-memory-access-based (DMA) communication, \tpm data is always fetched into CPU caches.
From there, it can be directly extracted via Foreshadow~\cite{Bulck2018} if the CPU is affected, or via Spectre~\cite{Kocher2018} indirectly if a suitable gadget exists in the enclave.
To prevent untrusted system software from interfering with the channel, we change the logic in the I/O controller hub. When a programmed I/O communication is initiated, the I/O controller hub does not process an incoming MMIO request from DMA. 
This is possible by distinguishing an I/O request from a DMA request in the \lstinline{LAD[3:0]} bus signal (\lstinline{00}: I/O read-write, \lstinline{10}: DMA read-write)~\cite{LPCbusSpec}. 
Thus, the system software can interact with the \tpm with a regular DMA request only when \sgx is not interacting with the \tpm. 
In our proof-of-concept implementation, we implemented this in QEMU's ICH9-I/O controller hub device, which can be updated through microcode. 

Although the size of CPU registers is limited, previous work showed that it is sufficient to compute AES~\cite{Muller2011tresor} and RSA~\cite{Garmany2013prime} entirely in registers. 
Cole and Prakash~\cite{Cole2020simplex} showed that in addition to general-purpose registers, sensitive data can also be stored in the Intel MPX \texttt{bnd} registers. 
As Intel MPX is not supported by Linux or GCC anymore~\cite{Phoronix2020MPX}, these registers can be used by an enclave without conflicting with any other existing software.

\noindent
\linebreak
\textbf{Interrupt Handling (\cheading{FC3}).}
On an interrupt, \sgx performs an \textit{Asynchronous Enclave Exit} (\lstinline{AEX}) to save the enclave execution state in the \textit{State Save Area} (SSA) before invoking the OS exception handling.
After the exception is handled, the enclave can resume by restoring the saved, protected state. 
While this is architecturally secure, RIDL~\cite{ridl2019}, ZombieLoad~\cite{Schwarz2019} and \AE PIC~\cite{Borrello2022AEPIC} showed that storing registers in the SSA leaves copies of the values in the line-fill buffer and load ports from where they can be leaked via these attacks, even if the enclave code is not vulnerable to side channels. 
Forcing \sgx to dump register contents to the SSA is always possible as an attacker can inject interrupts at arbitrary points during enclave execution~\cite{sgxstep2017}. 

To remedy this attack, \ourname does not allow the registers (\lstinline{BND0}-\lstinline{BND3}) holding potentially secret data to be saved directly to the SSA. 
In our proof-of-concept implementation, we encrypt the registers on an \lstinline{EEXIT}, \lstinline{EREMOVE}, or \lstinline{AEX} before storing them.
We use AES in counter mode, with the already-existing SGX sealing key as the encryption key and the number of asynchronous exits as the counter. 
Hence, even in the presence of transient-execution attacks, an attacker can only leak the encrypted secrets. 
Using the number of asynchronous exits as a counter has two advantages. 
First, an attacker has only one shot at leaking the (encrypted) secret. 
Second, an attacker cannot even detect if the secret has changed between two interrupts, a design issue that led to side-channel attacks on AMD SEV~\cite{Li2021cipherleaks}. 

As computations with secrets often require multiple general-purpose or SIMD registers~\cite{Muller2011tresor,Garmany2013prime}, it is also beneficial to prevent other registers from spilling secrets into the SSA. 
Similarly to protecting enclaves from traditional side-channel attacks, we see that responsibility with the enclave developer, as this can be entirely ensured in software. 
In contrast, without \ourname, a developer cannot write code such that secrets are not leaked via transient-execution attacks. 
If TSX is available, protecting intermediate results from spilling to the architectural domain is possible by relying on a compiler extension~\cite{Gruss2017Cloak}.
However, as TSX is deprecated, transient execution can be used as a (less-efficient) alternative, as shown in recent work~\cite{Wampler2019exspectre,Schwarz2019,Zhang2020exploring}.
Hence, we consider the actual computations with secrets out of scope for \ourname and only provide the means to securely store secrets and transfer them between \sgx and \tpm. 

\subsection{Porting SGX Functionality to TPM}
\label{sec:integration}

In this section, we demonstrate that given \ourname's secure channel, \sgx functionality can be transparently mapped to the \tpm using command authorization. 

\noindent
\linebreak
\textbf{TPM Command Mapping (\cheading{FC1}).}


Table~\ref{table:commMap} provides an overview of the mapping between TPM commands and the SGX commands and operations. 
Figure~\ref{fig:tpmconnle}.a shows the \ourname workflow for using the TPM as the backend for SGX-SDK functions handling keys. 
Other operations, such as reading a persistent counter from the \tpm follow the same idea. 
Typically, \sgx enclaves create keys via the \lstinline{sgx_get_key} command, which returns the actual key to the enclave. 
\ourname provides a modified version of that command (\circledbs{A.1}) with the same parameters to select the seeds for key derivation (e.g., \lstinline{MRENCLAVE}). 
The microcode gathers the calling enclave's identity from its \lstinline{SECS} (\sgx Enclave Control Structure, stores per-enclave metadata associated with each enclave) 
and uses it as seed parameters in the \lstinline{TPM2_CreatePrimary()} command (\circledbs{A.2}).
As with \lstinline{EGETKEY}, the TPM key is derived from the calling enclave's identity. 
In contrast to \sgx, \tpm returns a key handle (and optionally the public key and wrapped private key), which is then forwarded to the enclave for later use (\circledbs{A.3} and \circledbs{A.4}). 
If an enclave wants to sign or encrypt data with the \tpm key (\circledbs{B.1}), it has to use the corresponding key handle.
For persistent secure storage of the wrapped keys, an enclave can rely on the OS to store the data on the hard disk. 
Creating and using counters is similar to key handling. 
TPM counters are implemented in the \tpm's NVM.
Thus, creating a new counter equals creating a new dedicated NVM space with \lstinline{TPM_NVDefineSpace} and returning a handle to that space to the enclave. 
Via this handle, the enclave can read or increment the TPM-managed counter.
In contrast to the SGX counters, these counters cannot be reset~\cite{MateticAKDSGJC17} and are not limited to Windows~\cite{intelCounter}. 
For retrieving the time, TPM's \lstinline{GetTime} or \lstinline{Readclock} can be used. 
\tpm provides a secure clock signal with the granularity of 30~ns (LPC bus bandwidth is 33 MHz). 

\noindent
\linebreak
\textbf{SGX Key Generation with TPM.}
\sgx cryptographic keys are derived in the \lstinline{EGETKEY} instruction from different combinations of key derivation material~(KDM). 
The KDM consists of a list of enclave-related information and a list of CPU-known values.
The enclave information is \lstinline{MRSIGNER} (enclave developer's identity), \lstinline{MRENCLAVE} (hash of enclave memory pages), \lstinline{ISVSVN} (enclave security version number), \lstinline{ISVPRODID} (enclave product ID), \lstinline{KEYID} (a value supplied by the enclave), \lstinline{KEYNAME} (key type), and \lstinline{MASKEDATTRIBUTES} (enclave attributes).
This information is available in the \lstinline{SECS} created by \sgx for every enclave. 
The CPU-specific values are \lstinline{OWNEREPOCH} (platform owner identification), \lstinline{CPUSVN} (CPU version number), and \lstinline{SEAL_FUSES} (a secret known only to the processor).

A \tpm provides adequate secrets and functionality to achieve the same binding of keys as \sgx (\cf Figure~\ref{fig:tpmconnle}.b). 
\tpm offers \lstinline{TPM2_OWNERSHIP}, which can replace the \lstinline{OWEREPOCH}. 
Hence, if the platform changes the owner, all keys are invalided in the same fashion as for a changed CPU \lstinline{OWEREPOCH}. 
\lstinline{CPUSVN} is not a secret value but a counter denoting the current version of the CPU.
\sgx can share the \lstinline{CPUSVN} with the \tpm to be used as KDF input. \lstinline{SEAL_FUSES} binds all generated keys to the specific CPU. 
\tpm-generated keys can be bound to the specific \tpm through \tpm secret seeds (i.e., \lstinline{TPM2_CreatePrimary} command or \lstinline{TPM2_Create} for non-migratable keys). 
To bind generated keys in \ourname to both platform CPU and \tpm, \sgx sends a secret derived from \lstinline{SEAL_FUSES} to the \tpm as input to the \tpm key generation.
Lastly, it should be noted that \sgx uses an \sgx master derivation key (derived using \lstinline{Provisioning Secret}) as input to the KDF.
However, the purpose is undocumented~\cite{CostanD2016}. 
We suspect this has to do with \lstinline{AES-CMAC} used as PRF in \sgx.
Since \tpm uses SP800-108 as KDF with \lstinline{HMAC} as PRF, we do not address the master derivation key for now in our proof-of-concept implementation.


\noindent
\linebreak
\textbf{Enclave Authorization (\cheading{SC2}).}
To ensure that one enclave cannot have unauthorized access to another enclave's \tpm entities, \ourname uses \tpm's extended authorization policies (EAP). 
EAP policies are set during the creation of a \tpm entity, such as a key. 
Using an EAP-protected \tpm entity requires a successful TPM policy session that fulfills the entity's EAP. 
The CPU in \ourname dictates the EAP of newly created \tpm entities (using its role as \tpm driver for enclaves).
It handles the policy sessions with the \tpm, hereby supplying the necessary information for authorization from the KDM. 
With EAP, we can represent the same policies that are reflected in the KDM selection in default SGX.
For example, if a key is created with \lstinline{MRSIGNER} selected but not \lstinline{MRENCLAVE}, \ie it can be derived by all enclaves of the same developer, we represent this in an EAP that requires the enclave's \lstinline{MRSIGNER} value. 
When using the key, the CPU supplies the current enclave's \lstinline{MRSIGNER} value to the \tpm policy session.
Only if it matches the value set in the EAP at key creation time, that enclave can use the key.

\subsection{Limitations of the \ournameplain Implementation}
\label{appendix:securityanalysis}


Our proof-of-concept implementation demonstrates that \tpm and \sgx are very amenable for an integration and that this symbiosis is beneficial for enclave security (\cf \Cref{usecases} for case studies).
Our security discussion motivates further research into more securely integrating coprocessors with CPUs.
To create a secure channel between \tpm and CPU in our prototype, we depend on the chipset controller and firmware (\txt). 
For a production deployment, this dependency is not ideal. 
In the past, \txt was attacked several times~\cite{Wojtczuk2009,Sparks2008} through Secure Management Mode~(SMM), and Intel had to deploy patches~\cite{Wojtczuk2009} and implement SMM Transfer Mode (STM)~\cite{IntelSTM2019}.
STM is a sandboxed SMM handler using virtualization with VT-x and VT-d. 
However, it cannot be ruled out that further attacks against SMM or TXT are discovered.
Hence, adding SMM or TXT to the TCB is undesirable.

As the TPM is outside the CPU package, it is susceptible to \textit{cuckoo attacks}~\cite{Parno2008} when the client (here the CPU) is not assured of talking to the intended TPM chip. 
In that case, the connection to the TPM can be redirected to an attacker-controlled TPM. 
In our proof-of-concept implementation, the CPU uses an end-to-end encrypted channel with pre-shared keys to the TPM (\lstinline{TPM_TakeOwnership}). 
In this setting, the pre-shared key for the CPU is stored by the chipset and provided to the CPU during initialization. 
Thus, a cuckoo attack does not succeed as long as the TPM and the chipset have not been compromised. 
However, as argued above, being dependent on the chipset integrity is undesirable as it increases the TCB.

A coprocessor physically integrated into the CPU, such as Microsoft Pluton~\cite{msPluton}, can remove the dependency on the chipset for a secure, authenticated connection and is promising for securing enclave technology as it does not increase the TCB.
While we did not attempt such a tighter integration for the proof-of-concept in this paper, we provide functional objectives and requirements for a secure integration between a coprocessor and enclave that have to be fulfilled indepedently from how the coprocessor and CPU are connected.
Our prototype then demonstrates the resulting benefits at the concrete example of \tpm and \sgx.



	%
	
	\section{Extended Requirements Analysis}
	\label{app:ReAnalSystematization}

\begin{table}[t]
	\centering
	\setlength{\extrarowheight}{0pt}
	\addtolength{\extrarowheight}{\aboverulesep}
	\addtolength{\extrarowheight}{\belowrulesep}
	\setlength{\aboverulesep}{0pt}
	\setlength{\belowrulesep}{0pt}
	\arrayrulecolor{black}
	\adjustbox{max width=\hsize}{
	\begin{tabular}{llllllll|llll}
		\multicolumn{1}{c}{\multirow{2}{*}{}} &
		\multicolumn{7}{c}{\begin{turn}{0}\textbf{Coprocessors}\end{turn}} & 
		\multicolumn{3}{c}{\begin{turn}{0}\textbf{TEEs}\end{turn}} & \\
		\multicolumn{1}{c}{\textbf{Security Requirements}} & 
		\multicolumn{1}{c}{{\cellcolor[rgb]{0.753,0.753,0.753}}\begin{turn}{90}AMD~PSP\end{turn}} & \multicolumn{1}{c}{{\cellcolor[rgb]{0.753,0.753,0.753}}\begin{turn}{90}Intel PTT in Intel CSME \quad\end{turn}} & \multicolumn{1}{c}{{\cellcolor[rgb]{0.753,0.753,0.753}}\begin{turn}{90}Google Titan Chip ~~\end{turn}} & \multicolumn{1}{c}{{\cellcolor[rgb]{0.753,0.753,0.753}}\begin{turn}{90}Apple T2\end{turn}} &
		\multicolumn{1}{c}{{\cellcolor[rgb]{0.753,0.753,0.753}}\begin{turn}{90}Microsoft Pluton\end{turn}} & \multicolumn{1}{c}{{\cellcolor[rgb]{0.753,0.753,0.753}}\begin{turn}{90}Firmware TPM\end{turn}} & \multicolumn{1}{c|}{{\cellcolor[rgb]{0.753,0.753,0.753}}\begin{turn}{90}Hardware TPM\end{turn}} &
		\multicolumn{1}{c}{\begin{turn}{90}Intel SGX\end{turn}} &
		\multicolumn{1}{c}{\begin{turn}{90}AMD SEV\end{turn}} &
		\multicolumn{1}{c}{\begin{turn}{90}ARM TrustZone\end{turn}} &
		\multicolumn{1}{c}{\begin{turn}{90}Keystone~RISC-V\end{turn}} \\ 
		\hline
		
		\textbf{CC. Communication Channel}~ ~ & \multicolumn{1}{c}{} &  &  &  &  &  &  &  &  &  &  \\
		~ ~+ CC1.~~ Secure communication channel with CPU &  \yes & \scros & \scros & \scros & \unknown & \scros & \yes & \yes & \scros & \scros & \scros \\
		
	    ~ ~+ CC2.~ ~No dependencies on external buffers & \scros & \unknown & \yes & \yes & \unknown & \scros & \scros & \scros & \scros & \scros & \scros \\
		\hline
		
		\textbf{HC. Horizontal Access Control}~ ~ & \yes & \scros & \scros & \unknown & \yes & \yes & \yes & \yes & \yes & \yes & \yes \\
		\hline
		
		\textbf{VC. Vertical Access Control} ~ ~ & \scros & \scros  & \scros & \scros & \yes & \yes & \yes & \scros & \scros & \kinda & \scros \\
		
		\bottomrule \\
	\end{tabular}
	}
	\caption{Comparison of Coprocessors and Trusted Execution Environments (TEEs) against security requirements. (\yes = fulfilled; \kinda = partially fulfilled; \scros = not fulfilled; \unknown = Unknown)}
	\label{tab:systematizationRequirements}
\end{table}

 \begin{table}[t]
 	\centering
 	\setlength{\extrarowheight}{0pt}
 	\addtolength{\extrarowheight}{\aboverulesep}
 	\addtolength{\extrarowheight}{\belowrulesep}
 	\setlength{\aboverulesep}{0pt}
 	\setlength{\belowrulesep}{0pt}
 	\arrayrulecolor{black}
 	\adjustbox{max width=\hsize}{
 		\begin{tabular}{llllllla|alll}
 			\multicolumn{1}{c}{\multirow{2}{*}{}} &
 			\multicolumn{7}{c}{\begin{turn}{0}\textbf{Coprocessors}\end{turn}} &
 			\multicolumn{3}{c}{\begin{turn}{0}\textbf{TEEs}\end{turn}} & \\
 			\multicolumn{1}{c}{\textbf{Objectives}} &
 			\multicolumn{1}{c}{{\cellcolor[rgb]{0.753,0.753,0.753}}\begin{turn}{90}AMD~PSP\end{turn}} & \multicolumn{1}{c}{{\cellcolor[rgb]{0.753,0.753,0.753}}\begin{turn}{90}Intel PTT in Intel CSME \quad\end{turn}} & \multicolumn{1}{c}{{\cellcolor[rgb]{0.753,0.753,0.753}}\begin{turn}{90}Google Titan Chip ~~\end{turn}} & \multicolumn{1}{c}{{\cellcolor[rgb]{0.753,0.753,0.753}}\begin{turn}{90}Apple T2\end{turn}} &
 			\multicolumn{1}{c}{{\cellcolor[rgb]{0.753,0.753,0.753}}\begin{turn}{90}Microsoft Pluton\end{turn}} & \multicolumn{1}{c}{{\cellcolor[rgb]{0.753,0.753,0.753}}\begin{turn}{90}Firmware TPM\end{turn}} & \multicolumn{1}{c|}{{\cellcolor[rgb]{0.753,0.753,0.753}}\begin{turn}{90}Hardware TPM\end{turn}} &
 			\multicolumn{1}{c}{\begin{turn}{90}Intel SGX\end{turn}} &
 			\multicolumn{1}{c}{\begin{turn}{90}AMD SEV\end{turn}} &
 			\multicolumn{1}{c}{\begin{turn}{90}ARM TrustZone\end{turn}} &
 			\multicolumn{1}{c}{\begin{turn}{90}Keystone~RISC-V\end{turn}} \\ 
 			\hline
		
 		\textbf{SP. Security Properties}~ ~ & \multicolumn{1}{c}{} &  &  &  &  &  &  &  &  &  &  \\
 		~ ~+ SP1.~ ~~Strict hardware isolation & \yes & \yes & \yes & \yes & \yes & \scros & \yes & \scros & \scros & \scros & \scros \\
 		~ ~+ SP2.~ ~~Untrusted software dependency & \yes & \yes & \yes & \scros & \unknown & \scros & \yes & \scros & \scros & \scros & \kinda \\
 		\hline
		
 		\textbf{SSC. Secure Storage Capability}~ ~ & \multicolumn{1}{c}{} &  &  &  &  &  &  &  &  &  &  \\
 		~ ~+ SSC1.~~Confidentiality and Integrity of storage & \scros & \scros & \yes & \kinda & \kinda & \yes & \yes & \scros & \scros & \kinda & \scros \\
 		~ ~+ SSC2.~~Secure counter / Rollback protection & \scros & \scros & \kinda & \kinda & \kinda & \yes & \yes & \kinda & \kinda & \scros & \scros \\
 		\hline \\
		
%
%
		
 	\end{tabular}
 	}
 	\caption{Comparison of Security Coprocessors and Trusted execution environments (TEEs) against objectives.}
 	\label{tab:systematizationObjectives}
 \end{table}
%


In this section, we complement our requirements analysis from Section~\ref{ReAnalSystematization} and systematically compare how different solutions for secure coprocessors and TEEs match the requirements defined that Section, and, hence, which pairings of coprocessors and TEEs are most suitable for integration.
The considered coprocessors are AMD PSP~\cite{AMDPSPCCC2019}, Intel PTT running in Intel CSME~\cite{intelPTT2014}, Google Titan~\cite{titanM}, Apple T2~\cite{secureEnclaveAPPLE}, Microsoft Pluton, Firmware TPM, and Hardware TPM.
For TEEs, we consider Intel SGX~\cite{intelSGX}, AMD SEV, ARM TrustZone, and RISC-V Keystone.

\noindent
\linebreak
\textbf{Communication Channel (CC).} 
To briefly recap, for a secure integration between the security coprocessor and the application processor (AP), the communication channel between them must be secured from eavesdropping even in case of physical attacks, \eg bus sniffing (\textbf{CC1}), and there should not be any dependencies on buffers vulnerable to microarchitectural attacks that can leak sensitive data transferred via the channel (\textbf{CC2}).
From \Cref{tab:systematizationRequirements}, hardware \tpm fulfills CC1 (\textbf{Hardware TPM CC1:} \yes) by supporting end-to-end encrypting the communication with the TPM~\cite{TPM2Spec}. 
However, this channel does not avoid insecure buffers, and decrypted data at the CPU-side might still pass through such buffers. 
In comparison, AMD PSP also supports only CC1 (\textbf{PSP CC1:} \yes) as it is part of the application processor (AP) package, i.e., physical bus tapping is not feasible. 
However, secret data is transferred in plaintext and might pass through insecure buffers.
Google Titan and Apple T2~\cite{t2APPLE,secureEnclaveAPPLE} fulfill only CC2 (\textbf{Titan, Apple T2 CC2:} \yes) by not releasing secret keys outside their execution environment, i.e., they can never pass through insecure buffers.
However, secret data has to be exchanged over an unencrypted channel with the coprocessor for processing (e.g., decryption).
None of the TEEs is free of insecure buffers, as demonstrated by recent attacks.
Moreover, only SGX can inherently provide a way to establish a secure channel with a coprocessor to the best of our knowledge.
Though not inherent to SGX but to Intel CPUs, they support creating an end-to-end encrypted channel between TPM and CPU~\cite{Futral2013,intelTXT2021}. 

\noindent
\linebreak
\textbf{Horizontal Access Control (HC).}
TEEs can host multiple tenants.
For example, SGX supports multiple (parallel) enclaves.
We defined horizontal access control as a requirement to ensure that AP and coprocessor can distinguish between requests from mutually-untrusted tenants inside a TEE.
For instance, one enclave should not be able to access another enclave's data within the coprocessor.
A trusted entity, like AP, needs to create access or identity tokens that can identify the TEE tenants.
The tokens must be securely communicated to the coprocessor.
The coprocessor also needs an understanding of those tokens to control access to managed data and secrets.
AMD PSP can distinguish between requests originating from different SEV VMs (\textbf{PSP HC:} \yes).
Hardware \tpm and Firmware \tpm employ extended authorization policies (EAP) that can utilize such access tokens for access control to TPM-managed objects, like TPM-internal storage and keys (\textbf{Hardware \tpm, Firmware TPM HC:} \yes).
Microsoft Pluton is fully compliant with the TPM 2.0 specification~\cite{TPM2Spec} (\textbf{Pluton HC:} \yes).
Other coprocessors like Intel PTT, Google Titan, or Apple T2 do not intrinsically differentiate between their callers' identities or access rights.
All of the AP-based TEEs can fulfill this requirement because they can uniquely identify different enclave code that they host. 
They can provide this information on calls to the coprocessor (\textbf{SGX, SEV, TrustZone, Keystone HC:} \yes).
For example, in SGX, this would be the code measurement of the enclave by the CPU, and for ARM TrustZone the trusted OS in the secure world can identify the trusted applications.

\noindent
\linebreak
\textbf{Vertical Access Control (VC).}
As defined in Section~\ref{ReAnalSystematization}, both the AP-based TEE technologies and the coprocessor should support access control based on different security levels (\eg application, OS, or hardware) to prevent non-enclave code from being able to access enclave-owned entities in the coprocessor.
The access token to distinguish between different security levels needs to be generated and handled by a secure piece of code and be securely communicated to the coprocessor.
Hardware and firmware \tpm offer \emph{Locality} to distinguish between TPM commands originating from different security levels (\textbf{Hardware \tpm, Firmware \tpm VC:} \yes), but the locality of a command has to be communicated to the TPM by the CPU or firmware.
Microsoft Pluton is \tpm compliant in this regard (\textbf{Pluton VC:} \yes).
Other coprocessors like AMD PSP, Intel PTT, Google Titan, or Apple T2 do not inherently differentiate between the origin of commands.
Among the TEEs, ARM TrustZone CPUs communicate the current security level via a bit on the system-on-chip bus, but this bit has to be communicated on calls to the coprocessor (\textbf{TrustZone VC:} \kinda).
In contrast, SGX, SEV, and Keystone register when they are executing in enclave mode, but this security level is only used CPU-internally (\textbf{SGX, SEV, Keystone VC:} \scros). 

We additionally compare the TEEs and coprocessors regarding their security properties and provisioning of secure storage.

\noindent
\linebreak
\textbf{Security Properties (SP).}
Security properties are broken into \emph{Strict hardware isolation} and \emph{Untrusted software dependencies} (see \Cref{tab:systematizationObjectives}). 
\sgx uses PTT for certain trusted computing use-cases. PTT is housed inside the CSME~\cite{intelCSME2020} and connected through the DMI interface without any security around the communication channel. 
PTT uses the isolated execution environment powered by a 32-bit processor based on the Intel 486 architecture with a small dedicated SRAM (\textbf{PTT SP1: \yes}). 
CSME employs its own TCB OS with its own security ring, completely segregated from the platform security (\textbf{PTT SP2: \yes}).
However, the command buffer for PTT is configured by untrusted software such as the OS.
In \Cref{background:tpm}, we introduced Firmware and Hardware TPM and their features.
Hardware TPM is equipped with a low-end 32-bit coprocessor and has access to a small RAM (\textbf{Hardware TPM SP1: \yes}) and NV-storage. 
Google Titan~\cite{titanM} is designed and developed using a RISC-V processor combined with its own memory~\cite{titanM21} offering strict hardware isolation (\textbf{Titan SP1: \yes}). 
The goal of Titan with Android keymaster~\cite{AndroidKeystore} is to keep the key material secure with no possibility of extracting it from the hardware~\cite{AndroidKeystoreSecurityFeature}. 
The Titan backed keystore does not release keys to the system (\textbf{Titan SP2: \yes}).
Apple implements T2 with strict hardware isolation (\textbf{T2 SP1: \yes}). 
According to Microsoft, Pluton is fully compliant with the TCG TPM 2.0~\cite{TPM2Arch} specification. 
It uses a dedicated ARM M4 processor with \SI{128}{\kilo\byte} of \textit{Tightly Coupled Memory} (TCM) and \SI{64}{\kilo\byte} bootloader ROM \cite{AzureSphere}.
This configuration can provide an isolated hardware execution environment (\textbf{Pluton SP1: \yes}).
It is also unknown how the Pluton device is exposed to the platform (\eg as MMIO device) or if any untrusted software dependency is required (\eg preparing the command buffer) to perform actions on Pluton (\textbf{Pluton SP2: \unknown}).

\noindent
\linebreak
\textbf{Secure Storage (SSC).}
In \sgx, support for counters depends on the Platform Service Enclave and Intel ME, which often are not available in \sgx production deployments, and already deprecated~\cite{intelCounter} (\textbf{\sgx SSC2: \kinda}). 
Moreover, these counters can be simply reset by reinstalling the SGX platform software~\cite{MateticAKDSGJC17}. 
As \sgx stores counters inside the BIOS flash storage, they do not persist across system resets~\cite{MateticAKDSGJC17} (\textbf{\sgx SSC1:\scros}).
The NV-storage implemented inside the TPM chip offers complete confidentiality and integrity of storage of secrets (\textbf{Hardware TPM SSC1: \yes}).
The aforementioned storage can be used to implement secure counters which also can provide protection against version rollback attacks for both system and third party software, and hardware (\textbf{Hardware TPM SSC2: \yes}).
SEV SNP, \ie SEV with the secure nested paging extion, provides protection against firmware rollback but shows no evidence providing protection against virtual machine state roll back attack as it provides no secure storage capability (\textbf{SEV SSC2: \kinda} and \textbf{SEV SSC1: \scros}). 
Titan provides secure storage for the keys it generates or imports for its client (\textbf{Titan SSC1: \yes}). 
This storage can also store a secure counter for rollback protection.
However, we are unaware of existence of such counters (\textbf{Titan SSC2: \kinda}). 
TrustZone does not offer any secure storage facility. 
Manufacturer, \eg Microsoft, implement RPMB-based storage to store counters and fuses for other secrets such as device specific OEM keys. 
However, this storage is only available to system software (\textbf{TrustZone SSC1: \kinda}). 
While confidentiality and integrity of the T2 storage is given, third party code in the TEE cannot utilize it (\textbf{T2 SSC1: \yes}). 
T2 implements secure rollback protection only to the firmware, resulting in a limited counter support and rollback protection (\textbf{T2 SSC2: \yes}).
Pluton does not offer any secure storage facility other than programmable fuses to store platform specific cryptographic secrets (\textbf{Pluton SSC1: \kinda}). 
These fuses can be used to store counters to provide rollback protection. 
However, it is doubtful how much this protection can be extended beyond the platform (\textbf{Pluton SSC2: \kinda}).


	%
	\section{Other Platforms}
	\label{app:otherplatfrom}
	%

\ourname shows how a co-processor can be integrated with a TEE on x86. 
In Section \ref{harpocrates}, we present the integration process between a \tpm and Intel \sgx, and in Section \ref{usecases} we show the integration use cases. 
These use cases show that by integrating the functionalities of \tpm with Intel \sgx, security details, such as the launch key or quoting key, can be protected better. 
Moreover, \tpm provides crucial trusted building blocks (e.g., secure counter) to the TEE.


ARM platforms are not equipped with a \tpm. 
However, Google (with the Titan chip for Android) and Apple (with the Secure Enclave and T2 chip) equipped their ARM platforms with security co-processors to strengthen ARM TrustZone's security guarantees. 
However, these co-processors are still not deeply integrated into ARM TrustZone, with only limited use cases, \eg disk encryption, key generation, or attestation (see Section \ref{ReAnalSystematization}).
ARM TrustZone uses NS bit to distinguish between secure world and non-secure world, providing vertical access control. However, it lacks inherently a secure communication channel, which exists between Secure Enclave and T2. The UUID-based identification of Trusted Applets in TrustZone is by specification not a secure identity and cannot be used to implement horizontal access control. Moreover, \ourname uses the enclave code measurements as identities to control access to TPM objects. ARM TrustZone does not inherently use code measurements to identify secure code but would rely on the secure world OS for this task. Both Google Titan and Apple T2 do not support vertical access control (similar to TPM locality). Establishing secure identities for trusted applets for TrustZone and locality based access control for Google Titan and Apple T2 would be necessary steps towards fulfilling our requirements. 

RISC-V is a comparatively new platform and mostly limited to research prototypes. 
Specific RISC-V core implementations, such as the Boom core~\cite{CelioEECS2017} and Rocket core~\cite{AsanovicEECS2016}, allow implementing a special core within the processor called RoCC (Rocket chip coprocessor).  
RoCC accelerator can provide a secure register-based communication channel. RoCC can directly communicate to the physical core over registers, which can be used to establish a secure channel between keystone enclaves and RoCC. Only Keystone offers horizontal access control through enclave code measurement but not RoCC. RoCC and Keystone both do not support any type of vertical access control similar to \tpm locality. 
However, we believe RISC-V has more implementation efforts than fundamental limitations.
RoCC has proven its capability as an accelerator~\cite{Karandikar2021,Schmidt2015} but how much of a security coprocessor can be implemented in RoCC to support our requirements
is unclear.


	%
\end{subappendices}
\end{document}